\documentclass{aa}
\usepackage[varg]{txfonts}
\usepackage{natbib}
\usepackage{graphicx}
\usepackage{amsmath}
\usepackage{xcolor}
\usepackage[normalem]{ulem}
\bibpunct{(}{)}{;}{a}{}{,} 
\newcommand{\vprime}{\vec v\, '}

\def\dpar#1#2{\frac{\partial #1}{\partial #2}}
\begin{document}
\title{Stellar evolution models with overshooting based on 3-equation non-local theories}
\subtitle{I. Physical basis and the computation of the dissipation rate} 
\author{F.\ Kupka\inst{1,2,3}
\and F.\ Ahlborn\inst{4}
\and A.\ Weiss\inst{4}}
\institute{Dept. Applied Mathematics and Physics, Univ. of Applied Sciences, Technikum Wien, H\"ochst\"adtplatz 6, A-1200 Wien, Austria\\
email: \url{kupka@technikum-wien.at}
\and Wolfgang-Pauli-Institute c/o Faculty of Mathematics, University of Vienna, Oskar-Morgenstern-Platz 1, A-1090 Wien, Austria
\and Max-Planck-Institut f\"ur Sonnensystemforschung, Justus-von-Liebig-Weg 3, 37077 G\"ottingen, Germany
\and Max-Planck-Institut f\"ur Astrophysik, Karl-Schwarzschild-Stra{\ss}e 1, 85748 Garching, Germany
}
\date{Received xxx /
Accepted xxx }
\abstract
{Mixing by convective overshooting has long been suggested to play an important role for the amount of hydrogen available to nuclear burning in convective cores 
of stars. The best way to model this effect is still debated.}
{We suggest an improved model for the computation of the dissipation rate of turbulent kinetic energy which can be used in non-local models
of stellar convection and can readily be implemented and self-consistently used in 1D stellar evolution calculations.}
{We review the physics underlying various models to compute the dissipation rate of turbulent kinetic energy, $\epsilon$, in local and particularly in non-local
models of convection in stellar astrophysics. The different contributions to the dissipation rate and their dependence on local stratification and on 
non-local transport are analysed and a new method to account for at least some of these physical mechanisms is suggested.}
{We show how the new approach influences predictions of stellar models of intermediate-mass main-sequence stars
and how these changes differ from other modifications of the non-local 
convection model that focus on the ratio of horizontal to vertical
(turbulent) kinetic energy.} 
{The new model is shown to allow for a physically more complete description of convective overshooting and mixing in massive stars. Dissipation by buoyancy 
waves is found to be a key ingredient which has to be accounted for in non-local models of turbulent convection.}

\titlerunning{Turbulent convection for stellar evolution}
\authorrunning{F.\ Kupka et al.}
\maketitle
\section{Introduction}

From the early work of \cite{biermann32r} onwards research on convection has remained a major
challenge in stellar astrophysics, in particular as convection turned out to be one of the most important
mechanisms of energy transport and mixing in stars. As is described, for instance, in \cite{canuto09b}, when 
we compare the spatial scales of viscous processes derived from the results of \cite{chapman54b} on
fully ionised gases with the spatial scales of convective flow observed at stellar surfaces \citep{kupka17b}, 
stellar convection is characterised by very high Reynolds numbers. Stellar convective flows are thus
highly turbulent, even though the direct detection of turbulence is difficult due to the nature and resolution
of observational methods available to us (cf.\ \citealt{kupka17b}). 

To model this class of flows poses serious challenges for stellar structure and evolution models
(for introductions and reviews see, e.g., \citealt{cox04b}, \citealt{canuto09b}, \citealt{kupka17b}, \citealt{kupka20b}).
Due to the extreme range of scales in space and time numerical, hydrodynamical simulations
cannot be used directly in stellar evolution calculations (cf.\ the estimates given in \citealt{kupka17b}). 
Consequently, turbulent convection has to be modelled in a framework affordable for direct
coupling into one-dimensional stellar models of stellar evolution. The turbulent convection models (TCM) 
used in this approach differ widely in computational costs, physical completeness, and general principles 
considered in their derivation, from completely phenomenological to more systematic approaches based
on turbulence theory (see \citealt{kupka17b} for an overview). 

One methodological way to derive TCM equations which are suitable for stellar evolution calculations
is the Reynolds stress approach. The splitting of variables in turbulent flow into a mean and
a fluctuating component was first introduced by \cite{reynolds1894b}, followed by the suggestion
of \cite{keller25b} to consider this Reynolds splitting for a moment expansion approach that
was first completed by \cite{chou45b}. Dynamical variables such as velocity $\vec v$, 
density $\rho$, or entropy $s$, for example, can be subject to such splitting:
\begin{align*}
\vec v=\overline{\vec v}+\vprime, \,\,\,\, \rho=\overline{\rho}+\rho', \,\,\,\, s=\overline{s}+s', \,\,\dots
\end{align*}
Strictly speaking these are {\em ensemble averages} over different initial conditions. In practice,
the variables are also subject to spatial averaging, in one-dimensional stellar models typically over 
the $\theta$ and $\phi$ directions, to which the overbar in the above notation refers to whereas the component
with a prime refers to the fluctuating part of each quantity. 

Due to their immediate physical meaning the higher order combinations of the fluctuating parts
which appear in such Reynolds stress models of turbulent convection are also model predictions
of direct astrophysical interest. The second order moment of velocity fluctuations, 
characterising the {\em turbulent kinetic energy (TKE)} of the convective flow, is directly related to the highly 
efficient chemical mixing induced by convection. In stars with nuclear burning in convective 
cores this has a direct impact on the luminosity and the lifetime of
the nuclear burning phase. Similarly, the second order moment of velocity and entropy 
fluctuations, related to the convective flux, determines the energy transported by convection. 
Computing the convective flux allows predicting the temperature gradient in convective regions. Recently, 
the temperature gradient in core boundary layers of an intermediate-mass main-sequence star was 
probed using asteroseismology \citep{michielsen2021}, an observation that can directly be compared 
to results from a TCM.

Presently, the most commonly used theory to describe convection in stellar 
structure and evolution models is still the mixing length theory
\citep[][ MLT]{bv58b}. However, MLT is not able 
to describe the convective boundary in a physically accurate way. Observations
have shown that chemical mixing beyond the boundary of convectively
unstable regions, commonly known as \textit{overshooting}, is 
required \citep[see, for example,][]{maeder1981,bressan1981,pietrinferni2004}. In stellar
models using MLT parametrised ad~hoc mixing beyond the boundary is 
introduced to achieve this. Likewise, the temperature structure of an overshooting region 
cannot be predicted by MLT. These examples highlight the need for more physical theories of 
convection, like  TCM, being included in stellar structure and evolution models.

A large number of TCM have been developed 
\citep{xiong1997,canuto92b,canuto93b,canuto98b,li2001,li2007,kuhfuss86b,kuhfuss1987} 
which differ in the set of variables used  and the set of approximations and assumptions made (see \citealt{canuto93b}
and \citealt{kupka17b} for comparisons and a review). Among other physical effects the dissipation 
of TKE requires a careful discussion in the context of TCM. Acting as a sink term 
for TKE in overshooting layers the dissipation rate has a direct impact on the extent of convectively mixed regions. Assuming a 
Kolmogorov spectrum of turbulence the dissipation rate of TKE can conveniently be computed by a local 
expression involving  a dissipation length scale with a single constant parameter. This expression is, however,
inapplicable in non-local situations, encountered in layers adjacent to convectively unstable zones. To 
treat the dissipation of TKE in non-local convection models a physically more complete description of the 
dissipation rate is required \citep{zeman77b,canuto98b}.

We begin this paper by discussing local and non-local descriptions of the dissipation of TKE in Sect.~\ref{sect_diss}.  
From the dissipation rate in non-local convection theories we derive a model to account for the 
dissipation of TKE by buoyancy waves in overshooting layers in Sect.~\ref{Sect_GARSTEC}. In Sect.~\ref{sect_discussion} 
we then discuss implications of the improved dissipation model when applied to stellar models. For the computation of 
the stellar models we use the TCM derived by \cite{kuhfuss1987} implemented into the GARching STellar Evolution 
Code \citep[GARSTEC,][]{weiss2008}. The key assumptions and approximations of the \cite{kuhfuss1987} model are 
reviewed in Appendix~\ref{secKuh}. Using the local expression for the dissipation rate of the TKE we find an excessive 
overshooting extent beyond convective cores. When including the dissipation by buoyancy waves this overshooting 
is limited to a physically more reasonable range. This allows us to predict the convective core sizes and temperature 
structures of stars with different masses. We present our conclusions in Sect.~\ref{sect_conclusions}. A detailed discussion 
of the results obtained from the improved TCM can be found in \cite{paper2} (Paper~{\sc II} in the following).

\section{On the dissipation rate $\epsilon$ of turbulent kinetic energy}
\label{sect_diss}
The necessity to account for the dissipation rate of turbulent kinetic energy, 
$\epsilon$, in models of convection stems from the fact that it is not a negligibly 
small quantity. Indeed, the expression from which $\epsilon$ is computed is 
proportional to the kinematic viscosity $\nu$. The latter is small in stars compared
to the radiative diffusivity $\chi$ which results in the small values of the Prandtl 
number ${\rm Pr}=\nu/\chi$ typical for stars. Energy conservation requires 
$\epsilon$ to remain finite and non-negligible even in the limit of small viscosity 
(see \citealt{canuto97c}). Neglecting compressibility (for its modelling cf.\ \citealt{canuto97b}) 
we can compute $\epsilon$ from 
\begin{equation}  \label{eq_eps_spectrum}
\epsilon = 2\,\nu \overline{\left(\frac{\partial u_i}{\partial x_i}\right)^2} = 2\,\nu\,\Omega = 2\,\nu\,\int k^2E(k){\rm d}k
\end{equation}
in case of a locally isotropic, homogeneous flow. Here, $u_i$ is the \mbox{$i$-th} velocity component, 
$x_i$ is the \mbox{$i$-th} component of location, and $E(k)$ is the spectrum of turbulent kinetic 
energy as a function of wavenumber $k$.\footnote{Concerning notation the convention of 
                                                      summation over equally named indices is assumed.} 
Although convection is neither isotropic nor homogeneous on those large scales on 
which its contribution to energy transport is maximal, Eq.~(\ref{eq_eps_spectrum}) is 
a sufficient approximation to explain some basic properties of turbulent 
convection.\footnote{In real world systems the spectra
                 of turbulent kinetic energy, $E(k)$, usually depend on location ${\bf r}$ and in the most
                 general sense an averaging over directions in $k$-space would have to be performed,
                 i.e., $E(k)$ becomes a two-point correlation function $E({\bf k},{\bf r})$ and would
                 also have to account for density fluctuations.} 
In a quasi-stationary state where the amount of kinetic energy injected into the system 
per unit of time equals $\epsilon$, the enstrophy $\Omega$ of the flow increases, if $\nu$ decreases. 
The latter follows from the vorticity $\omega$ through
$2\,\Omega = \overline{\omega^2}$. Thus, $\epsilon$ is constrained by energy 
conservation and quantifies the amount of kinetic energy converted to thermal one.

If for a flow both the first and second Kolmogorov hypotheses hold
\citep{pope00b}, then there exists a range of length scales $\ell = \pi/k$
for which $\epsilon$ is independent of $\nu$ and independent of the details
of the large scale input of kinetic energy into the flow. This region
is known as the {\em Kolmogorov inertial range}. In that region $\epsilon$
is solely described by the exchange of energy between larger and
smaller scales. If this exchange peaks between neighbouring scales
(see \citealt{lesieur08b}), which is assumed to hold for turbulent flows
except for corrections due to intermittency (see also \citealt{pope00b}), 
it can be modelled as a flux in $k$-space. This is one of the basic inputs 
for the turbulence model of \citet{canuto96d} used in \citet{canuto98b} to 
justify the mathematical form and the constants involved in closure relations 
derived for their one-point closure Reynolds stress model of convection (see their Eq.~(9c)).

One important consequence for Eq.~(\ref{eq_eps_spectrum}) is the following 
one: if an inertial range exists, it can be shown to require 
\begin{equation}  \label{eq_Kolomogorov_spectrum}
   E(k)={\rm Ko}\,\epsilon^{2/3}\,k^{-5/3}
\end{equation}
to hold, i.e., a {\em Kolomogorov spectrum} to exist. Here, ${\rm Ko}$ is
the {\em Kolomogorov constant} which also turns out to equal $5/3$ in
the model of \citet{canuto96d} and \citet{canuto98b}, just as the {\em power 
law index} for $k$ in the spectrum Eq.~(\ref{eq_Kolomogorov_spectrum}).
Recalling  Eq.~(\ref{eq_eps_spectrum}) this demands that the contributions
of small scales $k$ to $\epsilon$ {\rm increase} with $k^{1/3}$ and a region
where the Kolmogorov inertial range no longer holds, just around the dissipation scale 
$k_{d}=\pi/{\ell}_d$, would have to be characterised more accurately than through 
Eq.~(\ref{eq_Kolomogorov_spectrum}) for a direct computation of $\epsilon$ from 
the spectral energy distribution $E(k)$ and Eq.~(\ref{eq_eps_spectrum}). Within 
a one-point closure model and thus in any of the prescriptions used in astrophysics 
to compute the convective flux inside a stellar structure code, this is not feasible 
and a different approach is required to compute $\epsilon$.

\subsection{Computation in local models: spectra and local limits}

One way around computing the spectrum $\epsilon(k)$ is to just compute its integral 
value $\epsilon$ from a model of $E(k)$ as follows. Assume that
$\nu$ is negligibly small. In the limit of vanishing $\nu$ the latter can ensure that
its product with $\int k^2\, E(k)\,{\rm d}k$ remains finite even though $\Omega$ might
increase indefinitely for arbitrarily large $k$. Hence, as in the derivation of 
Eq.~(5b) of \citet{canuto98b} and as also in their Section~6.4, assume that 
$E(k)$ is given by Eq.~(\ref{eq_Kolomogorov_spectrum}) from a certain 
value $k_0$ onwards, i.e., the entire energy spectrum is given by a Kolmogorov 
spectrum with an energy cutoff for $k < k_0$. Thus, $E(k)=0$ for $k < k_0$
and $E(k) \sim k^{-5/3}$ for arbitrarily large $k$ with $k \geqslant k_0$.
In this case, it is easy to first obtain $K$, the turbulent kinetic energy (TKE), 
from integration of $E(k)$ over all wavenumbers:
\begin{equation}  \label{eq_Kint}
   K = \int_0^{\infty} E(k)\,{\rm d}k = \int_{k_0}^{\infty} E(k)\,{\rm d}k, \quad {\rm if}\,\, E(k)=0\,\, {\rm for}\,\, k < k_0,
\end{equation}
and with Eq.~(\ref{eq_Kolomogorov_spectrum}) we obtain from Eq.~(\ref{eq_Kint}) that
\begin{equation}
   K = {\rm Ko}\,\epsilon^{2/3}\left.\frac{k^{-2/3}}{-2/3}\right|_{k_0}^{\infty} =
        \frac{3\,{\rm Ko}}{2} \epsilon^{2/3} k_0^{-2/3}.
\end{equation}
For $\ell_0 = \pi/k_0$ this can be quickly rearranged to yield
\begin{equation} \label{eq_MLT_epsilon}
   \epsilon = \pi \left(\frac{2}{3\,{\rm Ko}}\right)^{3/2} \frac{K^{3/2}}{\ell_0}  = c_{\epsilon} \frac{K^{3/2}}{\ell_0}=c_{\epsilon}\frac{K^{3/2}}{\Lambda},
\end{equation}
as shown in \citet{canuto98b}.\footnote{Note there is a typo in their Eq.~(5c) which should have the constant ${\rm Ko}$
                                              in the denominator.}
This is also the standard ``local'' or ``mixing length'' prescription for the computation of $\epsilon$.
It assumes maximum separation of the energy carrying scales around $\ell_0$ and the 
Kolmogorov dissipation scale $\ell_d$ (assumed to be negligibly small,
and not be confused with $\Lambda$).
Moreover, it assumes validity of the inertial range as if all the energy input were at one 
length scale only, i.e., at $\ell_0$, here set to be equal to $\Lambda$. 
All other scales for which $\ell \gtrsim \ell_0$ behave as if they were unaffected 
by the very small scales (scale separation) and also by the details of the energy input. 
Thus, a perfect energy cascade is assumed. Mixing length theory (MLT) {\em in addition}
replaces $E(k)$ with a $\delta$-function peaked at $l_0$ such that its integral yields 
Eq.~(\ref{eq_MLT_epsilon}). It is thus a ``one-eddy approximation'' where all the energy 
transport due to convection occurs on the critical (mixing) length scale $\Lambda$ 
which has to be computed for each layer. Either way, the challenge of computing $\epsilon$ turns 
into the challenge of prescribing $\Lambda$.

Evidently, this cannot be an accurate model, since at least a range of scales spanning
easily an order of magnitude (consider different granule sizes as an example) is expected
to contribute to energy transport at convective stellar surfaces such as those of our Sun.
Thus, Eq.~(\ref{eq_MLT_epsilon}) can at most be an estimate of order $O(1)$.
For some flows such as a shear flow in a pipe (Poiseuille flow), for which the mixing length 
formalism to compute the turbulent viscosity had originally been proposed by L.~Prandtl
(see \citealt{pope00b}, for example), this length can be fairly well constrained from geometrical 
arguments. Not surprisingly this is the application for which this prescription is most reliable.
For compressible convection on the other hand this length is much more difficult to constrain 
and the standard choice is to assume that
\begin{equation}  \label{eq_MLT_alpha}
  \Lambda = \alpha\,H_p
\end{equation}
where $\alpha$ is the MLT-parameter or mixing length parameter and $H_p$
is the local pressure scale height in the convective zone. This situation has motivated
\citet{canuto91b,canuto92c} to suggest a new convection model in which  
$\epsilon$ is computed directly from Eq.~(\ref{eq_Kint}). That removes the 
uncertainties introduced by the one-eddy approximation, but a scale length
$\Lambda$ is still introduced in this model. It compares the geometric size of
flow features which transport most of the energy with the length scales dominated 
by dissipation. This has permitted easy implemention into existing stellar evolution
codes based on MLT. The same approach was used by \citet{canuto96b}. 

But that concept collapses if an overshooting zone has to be 
modelled. In such a region, located just underneath or above a convectively unstable
zone, the convective flow is fundamentally non-local: the only way to sustain 
a non-vanishing solution is transport of kinetic and potential energy from the 
adjacent convective zone (cf.\ Sect.~10 in \citealt{canuto98b}). For such a region
there is no reason to assume that the prescription of Eq.~(\ref{eq_MLT_alpha}) 
with an $\alpha$ {\em independent of vertical location} can still hold.

Thus, even if other equations in a convection model are treated non-locally,
the continued use of Eq.~(\ref{eq_MLT_alpha}) with Eq.~(\ref{eq_MLT_epsilon})
along with a constant $\alpha$ even just within a single object may lead to inconsistent
or unphysical results, a fact long acknowledged in the atmospheric sciences 
by much more advanced modelling (see, for instance, \citealt{zeman77b}). 
As we show below, this is exactly the problem one encounters when using the 
3-equation Kuhfu{\ss} theory \citep{kuhfuss1987}, and it motivated the present work:
how to proceed and improve the computation of $\epsilon$ in such a case?

\subsection{Computation in non-local models: the dissipation rate equation}  \label{subsecdissip}

A common starting point for non-local models of convection is the 
dynamical equation for turbulent kinetic energy:
\begin{eqnarray}
 \lefteqn{\partial_t K + \partial_z \left(\frac{1}{2}\overline{q^2 w} + \overline{p w}\right)  =  
         g \alpha_{\rm v} \overline{w\theta} - \epsilon} \nonumber\\
        & & {} + \partial_z \left(\nu \partial_z K\right) + \frac{1}{2}C_{ii},   \label{eq_Kfull}
\end{eqnarray}
as given in \citet{canuto93b}, for example. In the Boussinesq case though, the
corrections due to compressibility given by the term $C_{ii}$ are zero. For the case 
of a low Prandtl number and if there are no contributions by a mean shear or rotation,
we obtain  \citep{canuto92b}
\begin{eqnarray}
\lefteqn{\partial_t K + \partial_z \left(\frac{1}{2}\overline{q^2 w} + \overline{p w}\right)  =  
        g \alpha_{\rm v} \overline{w\theta} - \epsilon} \label{eq_Keq}
\end{eqnarray}
which within the Boussinesq approximation is an {\rm exact} equation, though yet
unclosed. Here, $\partial_t$ and $\partial_z$
are partial derivatives with respect to time $t$ and vertical (radial) coordinate $z$. This is
a prognostic equation for the second order moment $K=\overline{q^2}/{2}$
with $\overline{q^2}=\overline{w^2+v_\theta^2+v_\phi^2}$ derived directly from the Boussinesq 
approximation of the Navier-Stokes equations through ensemble averaging. The non-local transport 
includes the flux of kinetic energy (in the Boussinesq approximation given by 
$F_{\rm kin}=\rho\,\overline{q^2\,w}/{2}$ with $w$ as the fluctuating component of vertical velocity) 
and of pressure fluctuations, $\overline{p\,w}$. This is to be balanced by local production,
$g \alpha_{\rm v} \overline{w\theta}$, and the local sink given by $-\epsilon$. Through
the cross-correlation $\overline{w\theta}$ the production is readily linked to the
convective (enthalpy) flux $F_{\rm conv} = c_p\,\rho\,\overline{w\theta}$. The latter
is exact in the Boussinesq approximation and can be generalised to a compressible flow. 
The quantities $g$, $\alpha_{\rm v}$, $c_p$, and $\rho$ are the local (vertical)
gravitational acceleration, the volume expansion coefficient, the specific heat at constant
pressure, and mass density.

To solve Eq.~(\ref{eq_Keq}) we need to know $\epsilon$.
The exact evolution equation for $\epsilon$ was first derived by \citet{davidov61b}. 
In their Sect.~3, \citet{hanjalic72b} emphasised\footnote{In the literature the model
      discussed here is known as $K-\epsilon$ model or ``Imperial College model'' since
      there the model had been developed by \citet{hanjalic72b}.}
why it is difficult to close this equation. But in the same paper they also point out 
how to proceed to derive a new equation which models the transport of $\epsilon$. 
One term (diffusional transport due to pressure fluctuations) is argued to be small on 
general grounds compared to other contributions while others are modelled such that 
the ensuing closure constants can be determined in the case of simple flows directly 
from experiments: decaying turbulence behind a grid and a constant-stress layer 
adjacent to a wall. Their model equation for $\epsilon$ eventually reads
\begin{equation}  \label{eq_epsilon_HL}
   \partial_t \epsilon + D_{\rm f}(\epsilon) =
     c_1 \epsilon K^{-1} P - c_2 \epsilon^2 K^{-1}      
       + \partial_z (\nu\partial_z \epsilon),
\end{equation}
where $P$ means production of dissipation (due to shear or buoyancy or both).
The term $\partial_z (\nu\partial_z \epsilon)$ is only relevant at moderate or low
Reynolds numbers and can always be neglected for small Prandtl numbers as is
the case for stars. The term $D_{\rm f}(\epsilon)$ was suggested to be parametrised as
\begin{equation}    \label{eq_diffeps}
   D_{\rm f}(\epsilon) \equiv \partial_z(\overline{\epsilon w})
   \approx -\frac{1}{2}\partial_z
   \left[(\nu_{\rm t})\partial_z \epsilon \right].
\end{equation}
where $\nu_t$ requires a model for turbulent viscosity such as\footnote{Note that this definition is different 
                      from \cite{canuto98b}, Eq.~(24c), which appears to have a typo.} 
$\nu_t = C_{\mu}\, K^2/\epsilon$ with a closure constant $C_{\mu}$. Although this term is mainly relevant for
moderate to low Reynolds numbers, it must be kept and modelled, since this is just what 
we also encounter in the case of overshooting zones. This is in contrast with terms only 
relevant for moderate to large Prandtl numbers (i.e., only in a non-stellar case) or which are small 
independently of the parameter space considered: those we can safely neglect for our applications. 
We emphasise that contrary to Eq.~(\ref{eq_Keq}) all contributions to Eq.~(\ref{eq_epsilon_HL})
contain closure approximations. Hence, Eq.~(\ref{eq_epsilon_HL}) is essentially a model for
$\partial_t \epsilon$ and not an exact evolution equation.

Eq.~(\ref{eq_epsilon_HL}) was reconsidered by \citet{canuto94b} and
\citet{canuto98b}, who also suggested the additional  
contribution to Eq.~(\ref{eq_epsilon_HL}) introduced in \citet{zeman77b}:
\begin{eqnarray}  \label{eq_epsilon}
   \partial_t \epsilon + D_{\rm f}(\epsilon) & = &
     c_1 \epsilon K^{-1} g \alpha_{\rm v} \overline{w\theta} - c_2 \epsilon^2 K^{-1}
       + c_3 \epsilon \tilde{N} 
       + \partial_z (\nu\partial_z \epsilon), \nonumber \\
                \tilde{N} & \equiv & \sqrt{g \alpha_{\rm v} |\beta|}.
\end{eqnarray}
Here, $\beta=-((\partial T/\partial z)-(\partial T/\partial z)_{\rm ad})$ is the superadiabatic gradient.
In addition to $c_1=1.44$ and $c_2=1.92$, which is close to the middle of the typical range of values in 
earlier work \citep{tennekes72b,hanjalic76b}, \citet{canuto98b} suggested $C_{\mu}=0.08$ from 
their turbulence model \citep{canuto96d}, which they obtained using Eq.~(\ref{eq_Kolomogorov_spectrum}). 

Before quantifying the new term $c_3\, \epsilon\, \tilde{N}$ more closely, the physical origin 
of the contributions to Eq.~(\ref{eq_epsilon}) requires some explanation. The first 
term on the right hand side provides a closure for the production of dissipation by buoyancy 
\citep{hanjalic72b}. The second term was discussed already in detail by \citet{hanjalic72b} 
and represents a closure for the combined effects of the exact terms describing the generation
of vorticity fluctuations through self-stretching in turbulent flows and the decay of turbulence due 
to viscosity. For the exact term of diffusion of $\epsilon$ by velocity fluctuations, $D_{\rm f}(\epsilon)$, 
both a down-gradient closure \citep{hanjalic72b} and a direct closure based on the flux of turbulent
kinetic energy \citep{canuto92b} have been proposed. The viscous diffusion term 
$\partial_z (\nu\partial_z \epsilon)$ is also part of the exact expression for diffusional transport and 
is suggested to be kept when modelling flows in the regime of low to moderately high Reynolds 
numbers, especially in the case of moderate to high Prandtl numbers (see \citealt{hanjalic76b}).

For buoyancy driven flows Eq.~(\ref{eq_epsilon_HL}) requires several changes in comparison
with \citet{hanjalic72b,hanjalic76b}. We refer the reader to the work by \citet{zeman76b} and 
\citet{zeman77b} which eventually allowed the derivation of Eq.~(\ref{eq_epsilon}). What follows 
from their and similar considerations is that, irrespectively of the detailed physical nature of 
increased local dissipation in the overshooting zone, a separately parametrised loss term that involves 
the superadiabatic temperature gradient $\beta$, or actually, the Brunt-V\"ais\"al\"a frequency, $\tilde{N}$, 
is needed. With hindsight gravity waves are expected to play the most important role as a source of $\epsilon$. 
As argued by \citet{zeman77b}, this involves a characteristic length scale which can be computed 
from the ratio of flow velocity $w^2$ and $\tilde{N}$.
It can also be viewed as the distance which eddies of a certain size that penetrate into the
stable layer with a certain lapse rate can travel until their potential energy is fully converted
into kinetic energy. It turns out that this yields the same expression as the parametrisation
of dissipation by internal gravity waves: their contributions may differ in magnitude, but
their functional form remains the same.

Hence, \citet{canuto94b} suggested that this term should indeed be added to the standard 
form of Eq.~(\ref{eq_epsilon_HL}). As they pointed out, this contribution also allows
to maintain stationarity in homogeneous, stratified turbulence as confirmed by
data from direct numerical simulations of shear turbulence by \citet{holt92b}. Thus,
\citet{canuto94b} suggested $c_3=0.3$ for stably stratified layers and  $c_3=0$ 
elsewhere to complete Eq.~(\ref{eq_epsilon}). \citet{canuto98b} followed that proposal. 

Clearly though, among all the parametrisations which appear in Eq.~(\ref{eq_epsilon}), 
$c_3\, \epsilon\, \tilde{N}$ remains the most uncertain one, but yet it is also crucial. Its 
choice requires to be tested carefully. Otherwise, the width of convective overshooting
may turn out to be sensitive to the detailed calibration of its
parameters. In Sect.~\ref{secEq11} we discuss more recent suggestions
to further improve the physical content of Eq.~(\ref{eq_epsilon}).

\section{A new model for the dissipation rate in non-local convection models in GARSTEC}   \label{Sect_GARSTEC}

\subsection{The problem: overshooting zones of convective cores growing unlimitedly during main-sequence stellar evolution}
\begin{figure}
	\centering
	\includegraphics{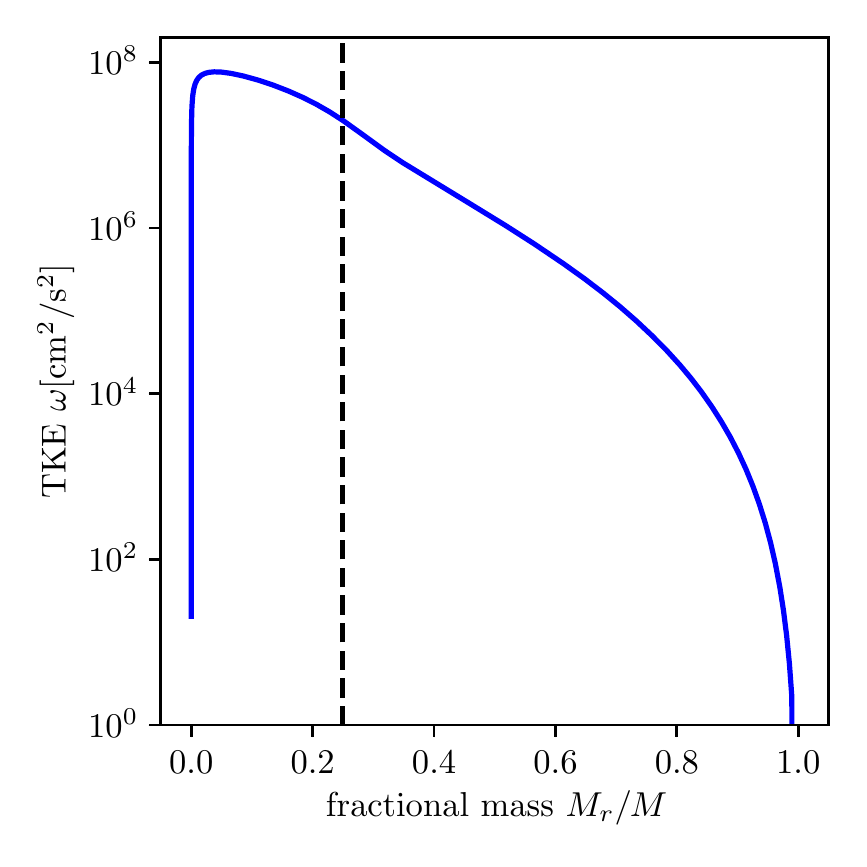}
	\caption{TKE as a function of the fractional mass for the original Kuhfu\ss~model. The formal Schwarzschild boundary, define by $\nabla_{\rm rad}=\nabla_{\rm ad}$, is indicated by a dashed black line.}
	\label{figenergyorig}
\end{figure}

The Garching Stellar Evolution Code (GARSTEC) (see \citealt{weiss2008}) offers several models 
to compute the contributions of convection to energy transport and mixing in stellar evolution 
calculations (including those of \citealt{bv58b}, \citealt{canuto91b}, \citealt{kuhfuss1987}). In particular, 
the model of \citet{kuhfuss1987} has been implemented (\citealt{Flaskamp02b}, \citealt{flaskamp2003})
in GARSTEC both in its 1-equation version, i.e., with an additional differential equation for 
turbulent kinetic energy, $K$, and in its full, 3-equation version
(\citealt{kuhfuss1987}; for a brief discussion of this model see App.~\ref{secKuh}).
The latter features differential equations for the TKE, the squared fluctuations of entropy, 
$\Phi$, and for the turbulent flux of entropy fluctuations, $\Pi$.   
Those three equations are essentially equivalent to the dynamical equations for the TKE, 
the squared fluctuations of temperature $\overline{\theta^2}$, and for the cross correlation 
between velocity and temperature fluctuations, denoted here by
$J=\overline{w\theta}$. The latter can be derived
from the phyically more complete model of \citet{canuto98b} by
assuming (i) an isotropic velocity distribution, (ii)
a local prescription to compute the distribution of the dissipation
rate $\epsilon$, (iii) the diffusion approximation
for the non-local fluxes, and (iv) some minor simplifications in the closures used in the dynamical 
equations.\footnote{Entropy gradients in turn are numerically easier to compute than the small differences 
 between temperature and adiabatic  temperature gradients during stellar evolution calculations.} 
As a variant, the 3-equation model may be used with local limit
expressions for the non-local transport terms for $\overline{\theta^2}$ as well as $J$. 
As a theoretical analysis shows (see \citealt{kupka20b} and references therein)
only a full 3-equation model can feature a countergradient or ``Deardorff'' layer where
$J$ is positive, while the superadiabatic gradient $\beta$ is
negative. Only in such a model both quantities 
can change their sign independently (the key to a positive convective flux in 
a countergradient stratification is the non-local transport of $\overline{\theta^2}$ 
as originally shown by \citealt{deardorff61b} and \citealt{deardorff66b}). However, 
in both the fully non-local and the local limit of the 3-equation model variant as described above, 
overshooting gradually mixes the entire star in a stellar evolution calculation for a 5 solar mass (B-type) 
main-sequence star. In Fig.~\ref{figenergyorig} we show the profile of the TKE as a function of fractional 
mass in this calculation. It can be seen that the energy extends substantially beyond the Schwarzschild 
boundary, reaching very close to the surface of the star. Due to the high efficiency of convective 
mixing the whole star would become essentially homogeneous which is
unrealistic, because the star would evolve from the hydrogen to the
helium main-sequence, i.e.\ to the left in the colour-magnitude
diagram, contrary to all observations \citep[see][ Chap.~23.1]{kippenhahn2012}. 
This problem was originally identified in the PhD thesis of \citet{flaskamp2003}.

To solve this problem \citet{flaskamp2003} suggested to give up the assumption of isotropy of 
TKE of the model of \citet{kuhfuss1987} in the overshooting (OV) zone and let the ratio 
of vertical to horizontal kinetic energy tend to zero. This limits the mixing efficiency in the outer layers 
of the OV zone, located above the stellar convective core, and avoids its unphysical growth throughout 
main-sequence evolution. If this simulation were plausible, also a more realistic model for the 
anisotropy of the convective velocity field, derived, for instance, from the stationary limit of 
Eq.~(19d) of \citet{canuto98b}, should solve this problem. Both variants of this approach 
are discussed below in Sect.~\ref{sect_flowaniso}.

\subsection{A comparison with a fully non-local Reynolds stress model}

\begin{figure*}
   \centering
   \includegraphics[width=0.4\textwidth]{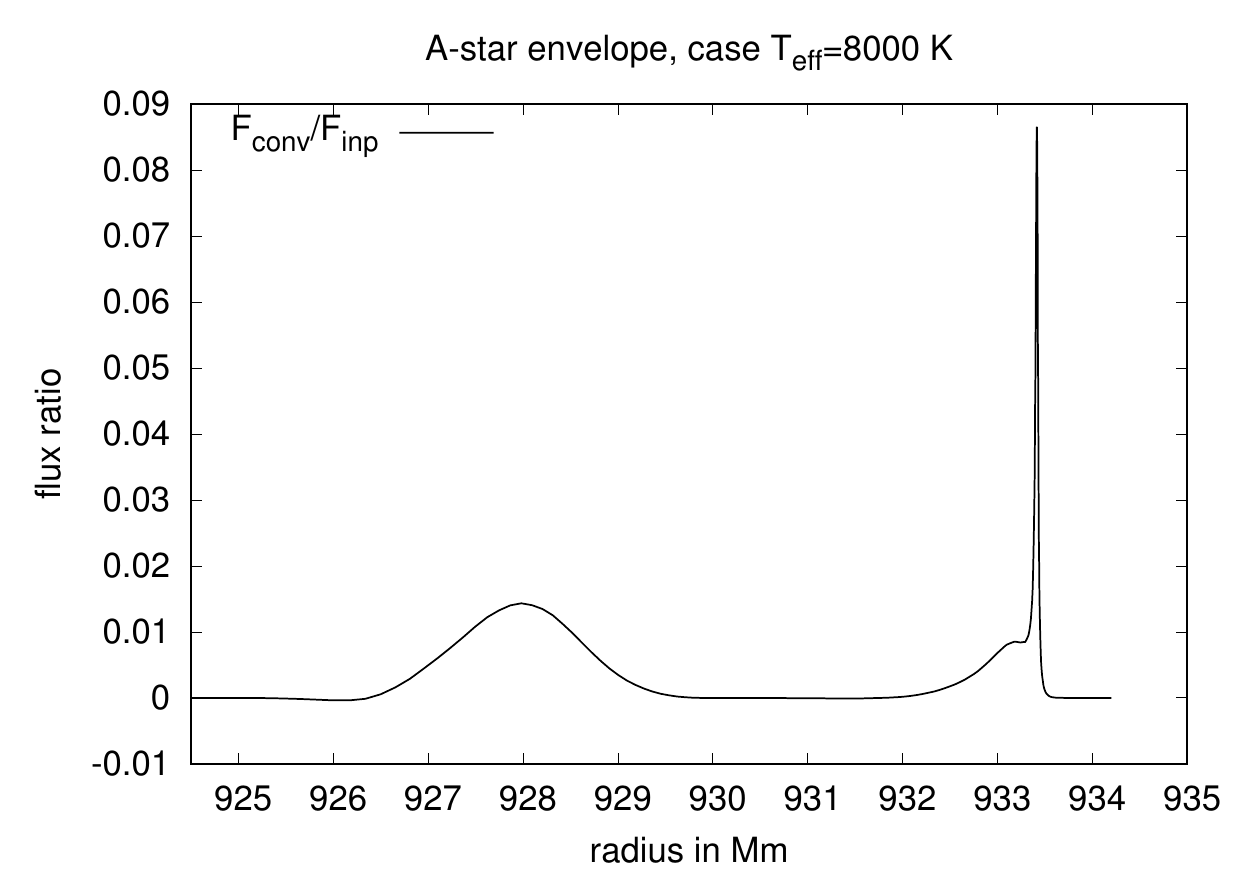}
   \includegraphics[width=0.4\textwidth]{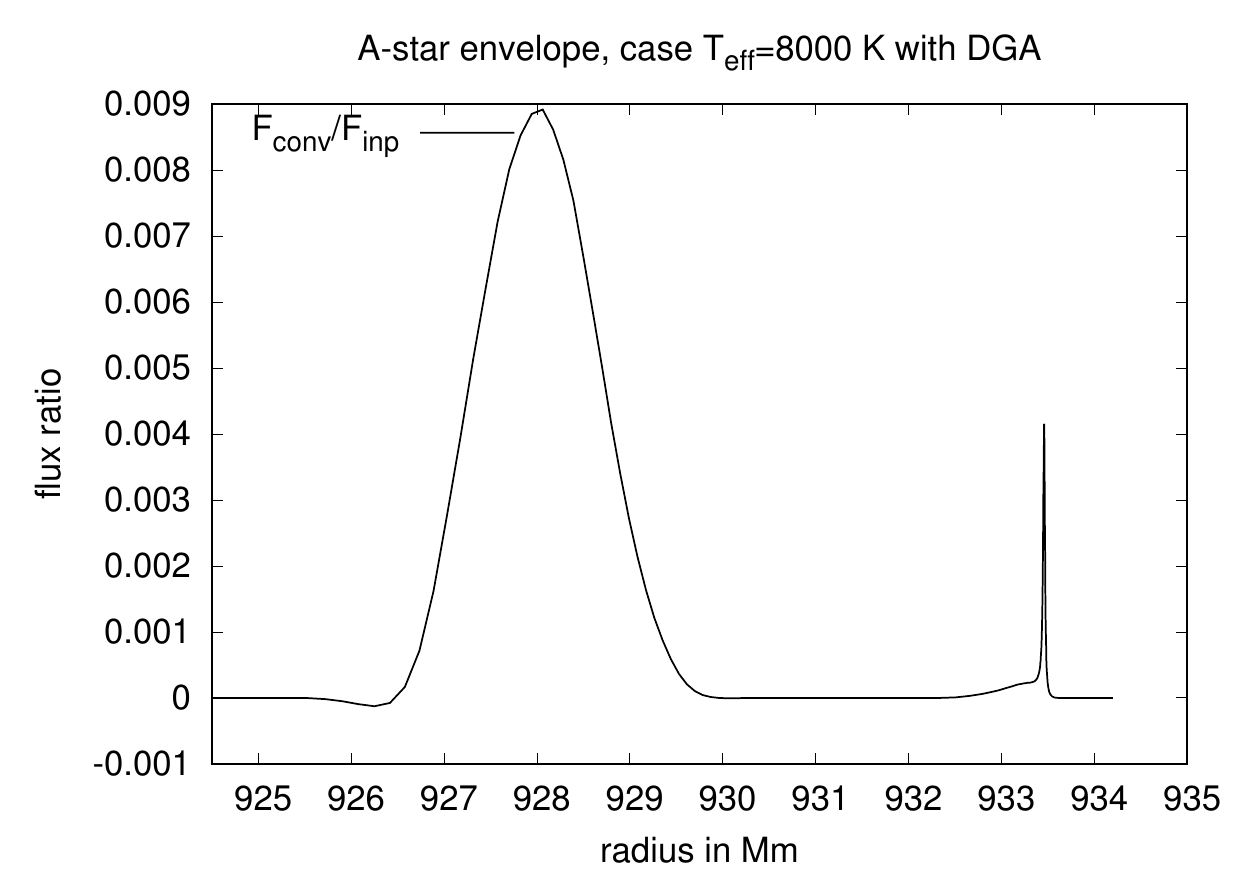}\\
   \includegraphics[width=0.4\textwidth]{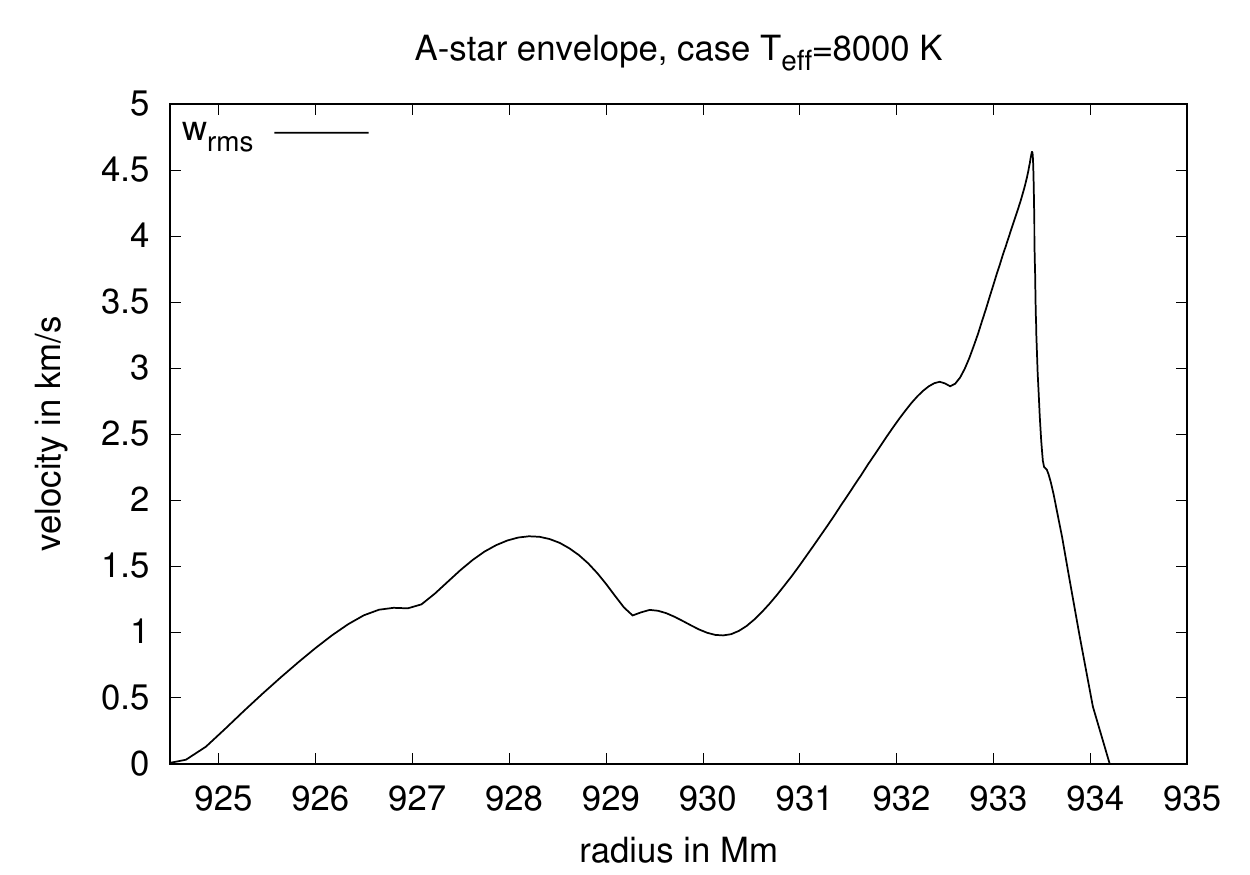}
   \includegraphics[width=0.4\textwidth]{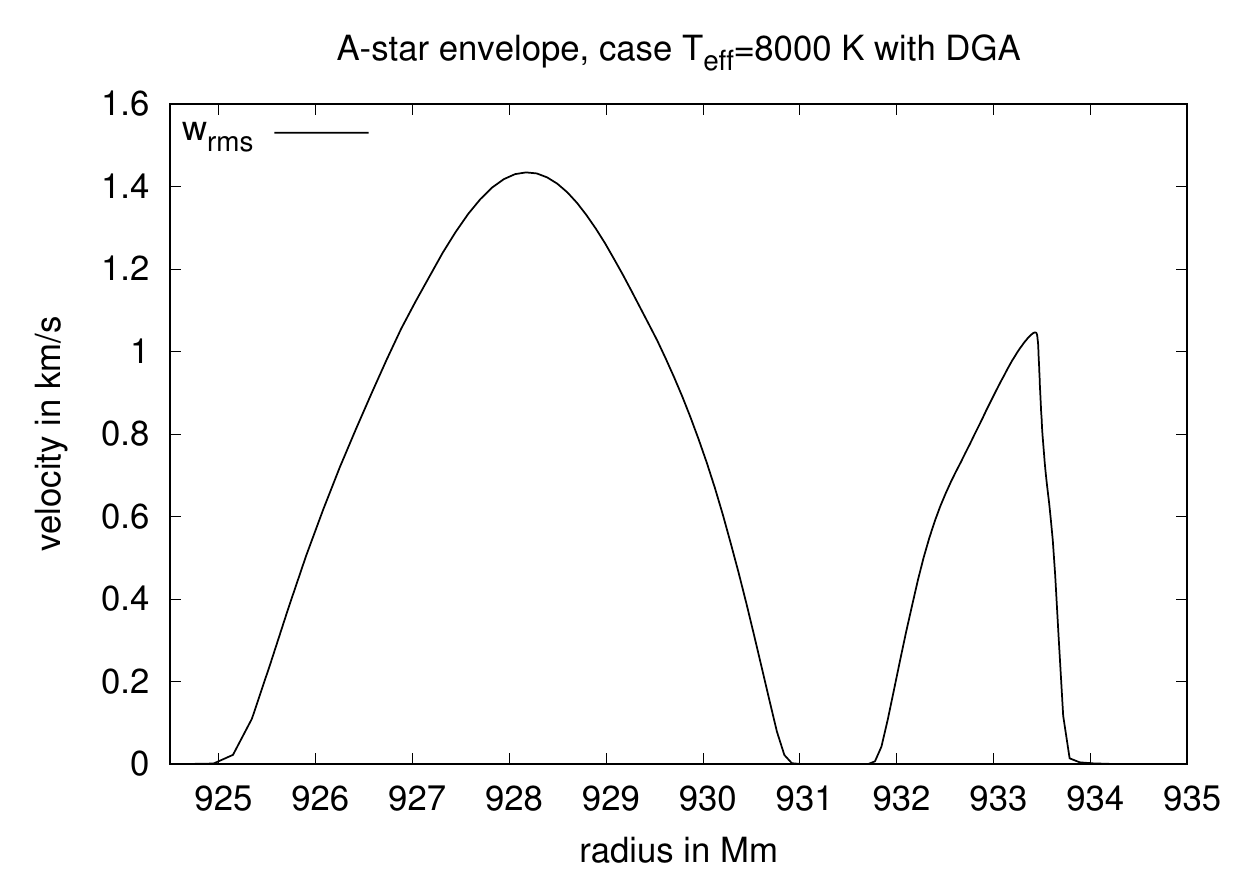}\\
   \includegraphics[width=0.4\textwidth]{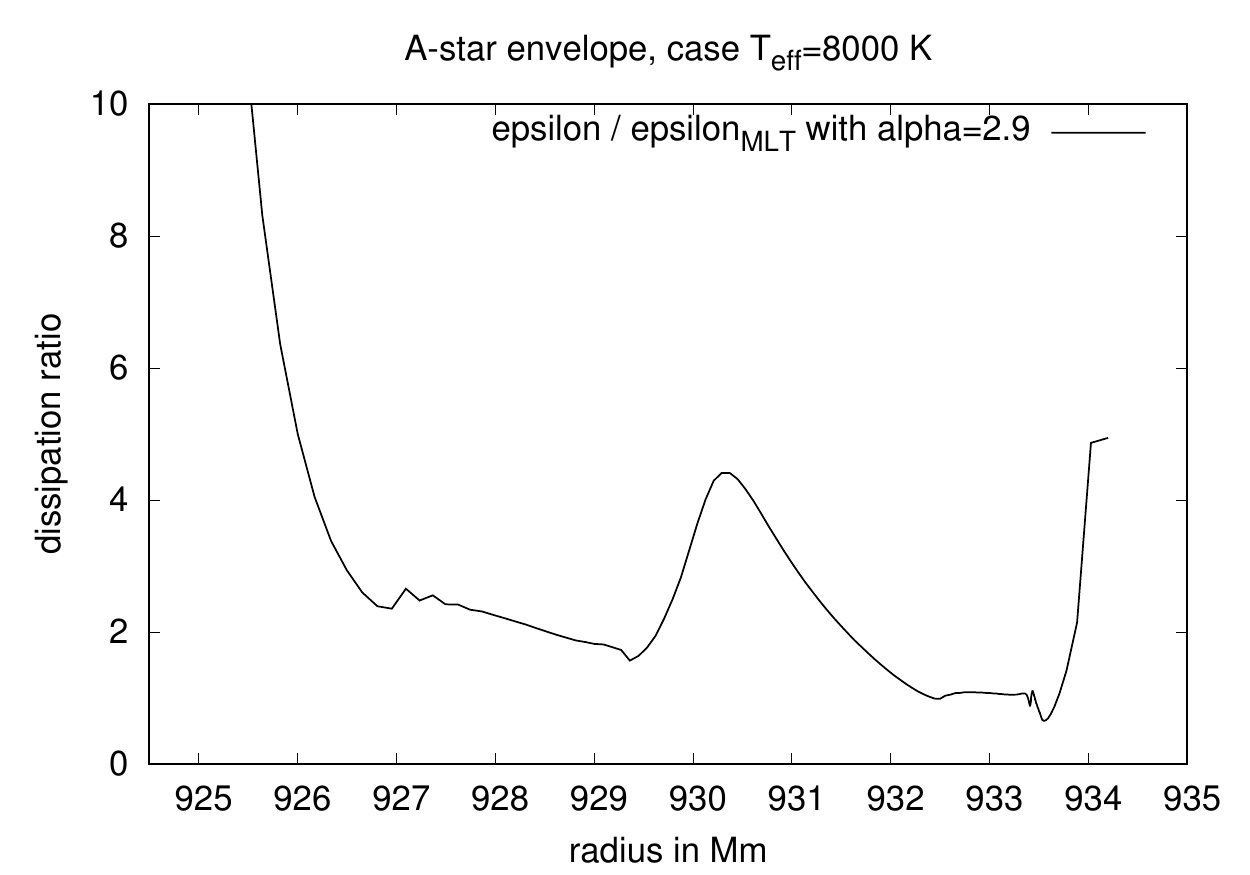}
   \includegraphics[width=0.4\textwidth]{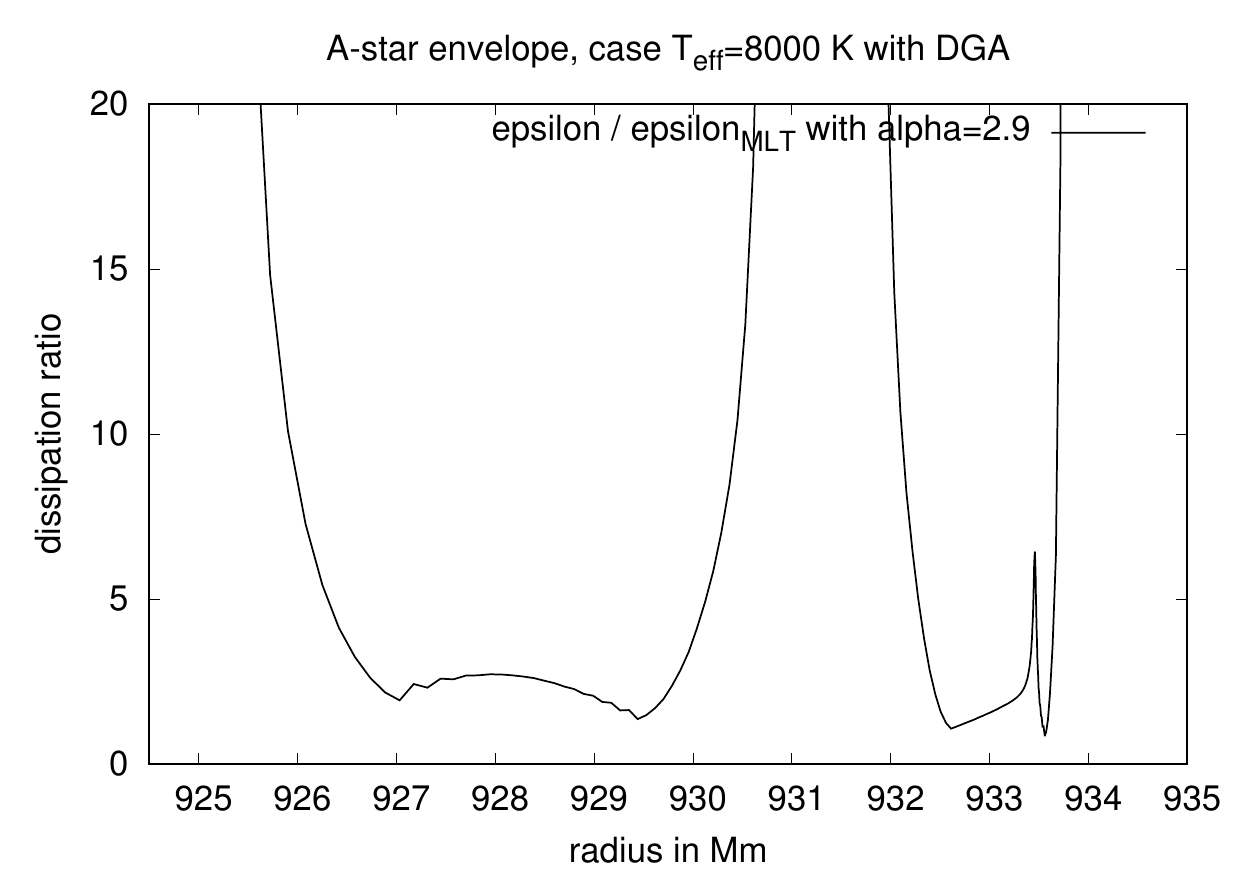}
      \caption{\textit{Left panels}: convective flux in units of total flux, root mean square vertical velocity in units of km/s,
           and dissipation rate $\epsilon$ from Eq.~(\ref{eq_epsilon}) relative to a value computed from 
           Eq.~(\ref{eq_MLT_epsilon}) and~(\ref{eq_MLT_alpha}) with $\alpha$ as given in the figure legend.
           \textit{Right panels}: same quantities as left panels, however, the downgradient approximation is used to compute third order
           moments instead of the full model used in \citet{kupka02b}. The results are for one of the A-star envelope
           models discussed in
           \citet{kupka02b}.
              }
         \label{fig1}
\end{figure*}

A progressive growth of the overshooting zone with time is not observed in 3D radiation hydrodynamical 
simulations of overshooting in DA white dwarfs \citep{kupka18b} either. Since the extension of the different zones 
in that case (Schwarzschild unstable convective zone with $J > 0$ and $\beta > 0$, countergradient region 
with $J > 0$ and $\beta < 0$, plume dominated region with  $J < 0$ and $\beta < 0$, and wave dominated 
region with $J \approx 0$ and $\beta < 0$) compare quite well with results from the non-local Reynolds stress 
model of \citet{canuto98b} solved in \citet{montgomery04b} for the same type of stars, the latter can 
provide a guideline for the behaviour of variables such as $\epsilon$ as a function of depth. The overall
structure of the OV zones and the behaviour of the convection related variables described in \citet{montgomery04b} 
is very similar to that one which had already been found for A-type main-sequence stars in \citet{kupka02b}
which in turn had been compared to earlier 2D RHD simulations of \citet{freytag96b}.

We hence use the Reynolds stress convection model calculations of \citet{kupka02b}
in Fig.~\ref{fig1} to illustrate the convective flux, the root mean square vertical velocity, and the 
dissipation rate as a function of depth. The left panels show results for the full third order moment model
while the right panels shows results computed using the downgradient approximation. For $T_{\rm eff} = 8000\,\rm K$ and $\log g$ slightly below 
the main sequence (see \citealt{kupka02b} for further details) we find two convective 
zones, an upper one due to ionisation of neutral hydrogen and a lower one caused by double-ionisation of 
helium. They are connected by an overshooting region at a radius of $\sim 931$~Mm and there is another 
overshooting region underneath the lower convective zone at $\sim 926.5$~Mm. 
For this setting we compare the computation of dissipation rates from the full equation of 
\citet{canuto98b} with the standard mixing length prescription for a range of bulk convective and overshooting layers.
Clearly, the dissipation rate $\epsilon$ becomes much larger than the value computed from the MLT prescription 
as soon as the plume region of the OV zones (with $J < 0$ and $\beta < 0$) is reached, and which can be determined 
from the behaviour of the convective flux. At the bottom of the lower overshooting zone, $\epsilon$ becomes 
even order(s) of magnitudes larger than the oversimplified MLT prescription would predict. 
Note that if the downgradient (diffusion) approximation is used to compute third
order moments such as $\overline{q^2w}$ in the model of \citet{canuto98b} (the non-local fluxes of
$K$, $J$, $\overline{\theta^2}$, and $\overline{w^2}$), a smaller 
overshooting is obtained in comparison with the complete third order moment model used in \citet{kupka02b}.
Hence, the two convection zones become separated at $T_{\rm eff} = 8000\,\rm K$ which allows observing 
this behaviour of $\epsilon$ even between the two convective zones. At lower $T_{\rm eff}$, for example 
at 7500~K, convection and overshooting are stronger also for the downgradient approximation of third order 
moments and the same behaviour is recovered as for the physically more complete third order moment model 
already for $T_{\rm eff} = 8000\,\rm K$. For that latter model the two convective zones become more tightly 
coupled and the increase of $\epsilon$ compared to the MLT prescription is eventually restricted to the lower  
overshooting zone only, for instance, for models with $T_{\rm eff} = 7200\,\rm K$. 

We hence can draw the following conclusions from solutions of the Reynolds stress model of
\citet{canuto98b} for convective envelopes of A-type stars:
irrespective of the various situations described above,
deep inside the plume-dominated region characterised by $J < 0$ and $\beta < 0$ the MLT prescription 
to compute $\epsilon$ begins to fail by entirely missing out the drastic increase in dissipation in that region.
However, the proper computation of $\epsilon$ is essential to determine
the extent of the mixed region, since it drains kinetic energy  
from the overshooting flow. From Eq.~(\ref{eq_MLT_epsilon}) one can immediately conclude that 
{\em underestimating} $\epsilon$ in the MLT framework can be easily caused by {\em overestimating} 
the mixing length $\Lambda$ or $\ell_0$.

\subsection{Reducing the mixing length in the OV zone}    \label{sect_reduced_ML}

There is also a physical argument why the mixing length must be limited and 
even gradually shrink in the OV zone on top of a stellar convective core. Taking 
$\Lambda$ to be about a pressure scale height at the convective core boundary 
results in a very large length scale. This is essentially the size of the convective core
itself. The claim that such a large structure penetrates into the radiative zone makes 
no sense, both from the viewpoint of available potential energy and from the viewpoint 
of the typical size of a convective structure. We note here that existing numerical
simulations of convective cores are actually for extremely different physical
parameter regimes, featuring mostly ${\rm Pr} \gtrsim 1$ or even ${\rm Pr} \gg 1$
(see, for instance, \citealt{Rogers13b}, \citealt{Rogers15b}, \citealt{Edelmann19b}).
They are unable to reproduce the very small levels of superadiabaticity
($\beta > 0$, but $|\beta/(\partial T/\partial r)_{\rm ad}|  \ll 1$) at realistic stellar
luminosities. This inevitably leads to excessive numerical heat diffusion and 
unrealistically small effective Peclet numbers (see \citealt{kupka17b} for a discussion). 
Numerical simulations of convective cores are hence likely also subject to the convective 
conundrum problem reported for the Sun (cf.\ \citealt{gizon12b}, \citealt{hanasoge16b}). Probably, 
they are not as reliable for guiding us as numerical simulations are in the case of convective overshooting 
near stellar surfaces (cf.\ \citealt{freytag96b}, \citealt{tremblay15b}, \citealt{kupka18b}, and many others). 
We return to the problem of comparing results on convective cores from stellar evolution models 
with 3D hydrodynamical simulations of convective cores in Sect.~\ref{sect_discussion_modsim}.
In the following, we thus use a different chain of arguments to derive an improved estimate of $\Lambda$.

\begin{figure}
	\includegraphics{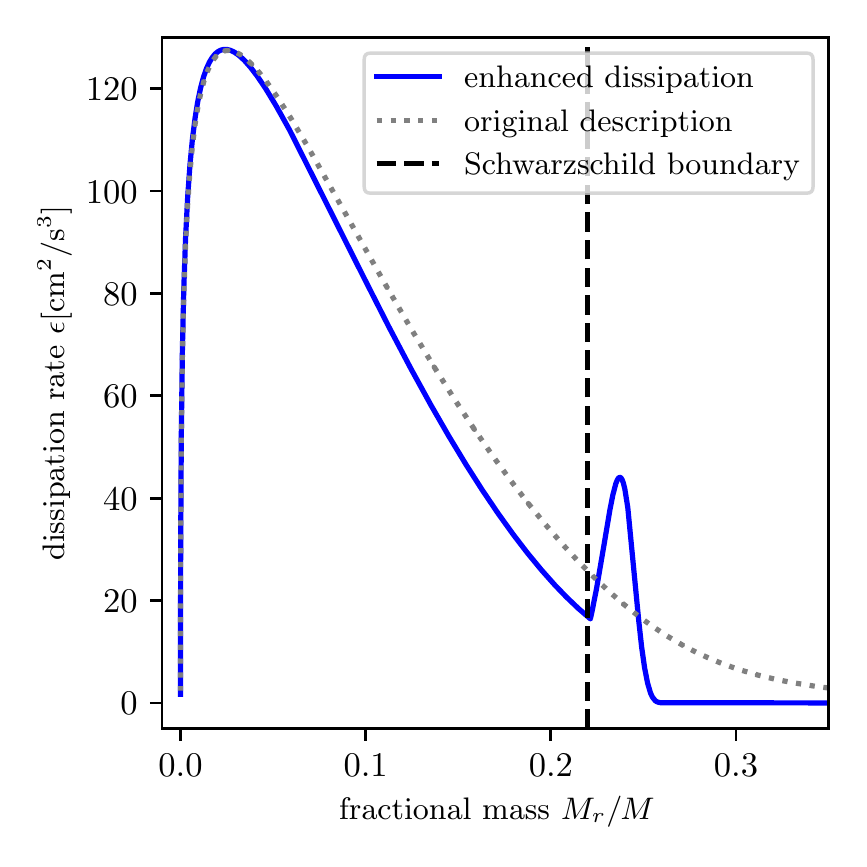}
	\caption{Dissipation rate as a function of fractional mass for the original Kuhfu\ss~model and the 
		     Kuhfu\ss~model including an ad hoc exponential decay of the dissipation length, 
		     shown with a grey dotted and a blue continuous line, respectively. The ad~hoc exponential decay of 
		     the dissipation length leads to an increased dissipation rate at the beginning of the overshooting zone, 
		     indicated by the local maximum beyond the Schwarzschild boundary,
		     followed by a sharp drop due to the rapid decay of TKE.
		     The models have been chosen to have the same
                     maximum TKE.}
	\label{figdissexp}
\end{figure}

As a very first step, one could let $\Lambda$ decay to zero within the OV zone, 
either linearly or  exponentially, from the value it has at the boundary of the convective zone.
This ad~hoc ``fix'' has been implemented into GARSTEC. The exponential decay model was chosen 
and {\em indeed} this easily stops the growth of the overshooting zone as a function of stellar evolution
time. The so enhanced dissipation rate introduced can be seen in Fig.~\ref{figdissexp} at the outer edge 
of the convective region. The model including the exponential decay has a central hydrogen abundance of 
0.6. The stellar model computed with the original Kuhfu\ss~model was chosen to have the same maximum TKE in 
the convection zone to make the dissipation rates comparable.

Physically plausible extensions of the OV zone can be obtained from a ``reduction factor'', which forces 
an e-folding extent of the ``decay'' of the mixing length of 2\% to 6\% of the mass of the Schwarzschild-unstable region.
In a $5\,M_\odot$ main-sequence star this limits the OV zone to contain  about 12\% to 29\% in terms of the Schwarzschild
core mass. The relative extent of the overshooting region in terms of the Schwarzschild core mass remains mostly constant 
along the main-sequence. For an e-folding extent of 4\% the overshooting region contains about 5\% of the stellar mass at 
the beginning of the main-sequence while it is shrinking to about 2\% of the total mass at the end of the main-sequence.
The procedure introduces a free parameter, but it is sufficient as a proof of concept: a physically more complete model of 
$\epsilon$ constrains the OV contrary to earlier, alternative explanations that require unphysical parameter values to do so 
(such as $\overline{w^2}/ K \rightarrow 0$ 
which is at variance with \citealt{kupka18b}, see Sect.~\ref{sect_flowaniso} below).

\subsection{Boundary conditions and regularity constraints}

As a prerequisite to derive an improved estimate for $\Lambda$ we first discuss its
asymptotic behaviour in the centre of a convective core.
Regularity properties of non-local models of convection at the centre of 
stellar cores are a rather delicate issue which has been analysed in \citet{roxburgh07b}. 
Under the assumption that non-zero convective motions 
can also occur at the centre of a convective core, for the second order moments they demonstrated 
that $\overline{w^2}$, $K$, and $\overline{\theta^2}$ are all positive and have an even order
expansion in $r$ just like the gas pressure $P$. Moreover, from their Eq.~(11), the horizontal
component of TKE has to balance the vertical component in the sense that 
$\overline{v_r^2} = \overline{v_{\theta}^2} = \overline{v_{\phi}^2}$ for the velocity components
in spherical polar coordinates $(r,\theta,\phi)$. Hence, $\overline{w^2}/K = 2/3$ and the flow
is isotropic. Clearly, also $\epsilon$ has to be positive in this case. 

Thus, if the relation $\epsilon = c_{\epsilon} K^{3/2} / \Lambda$ is used, a positive
$\Lambda$ guarantees positivity of $\epsilon$. An appropriate prescription which ensures 
this property is to use the curvature of the pressure profile to define a local scale height, 
since its gradient vanishes at the centre.
This has been worked out in \citet{roxburgh07c} where the scale height at the centre is defined from
$H_c^2 = -P / (\partial^2 P/\partial r^2) = 2 r H_p = 3 P_c / (2\pi G \rho_c^2)$ and the subscript
$c$ denotes the value of the local scale height, $H_c$, of pressure, $P_c$, and density, $\rho_c$,
at the centre (and $G$ is, of course, the gravitational constant). For the centre, $\Lambda = \alpha H_c$
and in general $\Lambda = \alpha \min(H_p,H_c)$. \citet{roxburgh07c} suggest a smooth interpolation
between the limit at the core centre and the expression for $H_p \gg
H_c$.

For reasons of regularity and energy conservation, $F_{\rm conv} \rightarrow 0$ at the centre
in that case, which is fulfilled by the above prescription of the mixing length. An  
expansion in odd powers of $r$ is found for the Reynolds stress equation 
for $J$ and thus for $F_{\rm conv}$ \citep{roxburgh07b}. This implies non-trivial constraints
on closures for the third order moments. \citet{roxburgh07b} demonstrate that the downgradient
closure forces the core centre to be convectively neutral ($J \propto r^3$ instead of $J \propto r$)
while other closures have to be modified to ensure regularity of the solution.

In GARSTEC, the \cite{wuchterl1995} prescription for $\Lambda$ 
is used by default. This requires a different approach at the core centre, as it assumes 
$\Lambda \rightarrow 0$ for $r \rightarrow 0$. Thus, in GARSTEC, it is ensured by power series 
expansions that the convective variables are not forced to zero while the temperature gradient 
at the centre is the adiabatic one. As a result, the convective quantities become small in the 
central region (see Paper~{\sc II}). Are differences between these two rather different prescriptions 
for the convective variables in the centre of a convective core relevant for applications? Fortunately,
it turns out that they remain constrained to less than the innermost 10\% of the stellar core. In either 
case, the stellar core is predicted to be fully mixed and has a temperature gradient close to the 
adiabatic one. For this study, we hence prefer to stay within the standard setup used for GARSTEC,
i.e., the prescription for $\Lambda$ proposed by \cite{wuchterl1995}.

\subsection{Some input from the dissipation rate equation}
Can we carry over some of the physics contained in Eq.~(\ref{eq_epsilon_HL}) or  Eq.~(\ref{eq_epsilon})
into a local model for $\epsilon$ which avoids the solution of an additional differential equation? 
If we model the non-local transport of TKE in Eq.~(\ref{eq_Keq}) by a downgradient approximation, 
the closure $\overline{w\epsilon} = (3/2)\tau^{-1}\,F_{\rm kin}$ relates $\overline{w\epsilon}$ to $\partial_z \overline{w^2}$ in Eq.~(\ref{eq_epsilon}).
The same behaviour is found for a direct downgradient closure for $\overline{w\epsilon}$ (i.e., computing it from $\partial_z \epsilon$)
as for example Eq.~(\ref{eq_diffeps}). Let us hence assume a local
approximation for $D_f(\epsilon)$, the non-local flux of $\epsilon$,
which replaces the derivatives of the outer divergence operator and the gradient 
operator in Eq.~(\ref{eq_epsilon}) by a product of reciprocal length scales, $1 / {\ell}^2$. 
Inspecting Eq.~(\ref{eq_epsilon}), for the sake of simplicity, it appears desirable to model as many 
contributions as possible by expressions of type $\epsilon^2/K \propto \epsilon / \tau$. Instead 
of a diffusion length scale ($\ell$) we hence use the characteristic transport time scale $\tau = 2K/\epsilon$ 
to approximate $D_f(\epsilon) \propto -\alpha_{\epsilon} \epsilon / \tau$. 
The same can be done also in the case of Eq.~(\ref{eq_epsilon_HL}).
If we furthermore assume 
the local limit of Eq.~(\ref{eq_Keq}), $P = P_{\rm b} = \epsilon$, i.e.\ production of TKE by 
buoyancy equals its dissipation, and if we also assume $c_3 = 0$, we obtain the following 
approximation for both Eq.~(\ref{eq_epsilon_HL}) and  Eq.~(\ref{eq_epsilon}):
\begin{equation}   \label{eq_eps_alpha}
  -\alpha_{\epsilon} \epsilon / \tau = 2\, c_1 g \alpha J / \tau - 2\, c_2 \epsilon / \tau.
\end{equation}
To remain consistent with $g \alpha J = \epsilon$ we have to require that 
$\alpha_{\epsilon} = 2 c_2 - 2 c_1$ if $\epsilon$ itself is computed from 
Eq.~(\ref{eq_MLT_epsilon})--(\ref{eq_MLT_alpha}). In this case we obtain a completely 
local model for the computation of $\epsilon$.

We can use Eq.~(\ref{eq_eps_alpha}) to understand some implications from the different 
physical contributions which its physically more complete counterpart, Eq.~(\ref{eq_epsilon_HL}), 
would instead account for. To this end let us relax the requirement $P_{\rm b} = \epsilon$ in Eq.~(\ref{eq_Keq})
somewhat. In this case, whether the 1-equation or the 3-equation version of the \citet{kuhfuss1987} model 
is used (cf.\ Appendix~\ref{secKuh}), due to the non-locality of the flux of kinetic energy
in Eq.~(\ref{eq_Keq}), $\partial_z(\overline{q^2 w}/2) \neq 0$, there is always a point where $J=0$
(cf.\ Chap.~5 in \citealt{kupka20b}). At such a point, $\alpha_{\epsilon}=2 c_2$ is required from 
Eq.~(\ref{eq_eps_alpha}) for a non-vanishing dissipation rate $\epsilon$. Right next to 
such a point, where $\epsilon > 0$ with $J<0$, a value of $\alpha_{\epsilon} > 2 c_2$ would be 
required whereas $\alpha_{\epsilon} < 2 c_2$ where $J>0$. So $\alpha_{\epsilon}$ would have 
to be a function that has to be fine-tuned to obtain consistent results from Eq.~(\ref{eq_eps_alpha}) 
in the vicinity of $J=0$. Moreover, because of the downgradient closure for $\overline{w\epsilon}$
also constraints on $\overline{w^2}/K$ would be imposed.

Such constraints appear unphysical: Eq.~(\ref{eq_eps_alpha}) does not 
provide a good starting point for a local model capable to capture at least the main gist of  
either Eq.~(\ref{eq_epsilon_HL}) or  Eq.~(\ref{eq_epsilon}). To proceed we need a physically more
complete model for $\epsilon$, i.e., we either have to abandon the mixing length prescription altogether 
or we need a more complete model equation than Eq.~(\ref{eq_epsilon_HL}) to start from. Let us hence
first have a look at Eq.~(\ref{eq_epsilon}), i.e., we no longer impose $c_3=0$ everywhere. The sibling 
of Eq.~(\ref{eq_eps_alpha}) which accounts for the production of dissipation by gravity waves in stably 
stratified fluid then reads:
\begin{equation}   \label{eq_eps_alpha_NBV}
  -\alpha_{\epsilon} \epsilon / \tau = 2\, c_1 g \alpha J / \tau - 2\, c_2 \epsilon / \tau + c_3 \epsilon \tilde{N}.
\end{equation}
If we were to combine this equation with the 1-equation model of \citet{kuhfuss86b}, $\beta$ and $J$ change 
sign at the same point so the perfect balancing constraint between $D_f(\epsilon)$ and  $-2\, c_2 \epsilon / \tau$
reappears. In the region where $J < 0$, more freedom of how $D_f(\epsilon)$ behaves is permitted.
This changes once we switch to the 3-equation model of \citet{kuhfuss1987}: since $\beta$ and $J$ then
change sign at different locations, $\alpha_{\epsilon}$ is no longer forced by $c_2$ at any point.
In the end, the $c_3 \epsilon \tilde{N}$ contribution decouples both $D_f(\epsilon)$ and $\overline{w^2}/K$
from peculiar constraints required to be fulfilled at where $\beta=0$ or where $J=0$.

On the other hand, now there is an efficient local source for $\epsilon$ also where $\beta < 0$. This
is particularly important for the 3-equation model which through its countergradient layer permits
much larger enthalpy (and hence also TKE) fluxes in this region: considering that property it is 
understandable that 
the 3-equation model can be prone to large overshooting,
unless the latter is limited by efficient dissipation. And this is just what gravity waves can provide.

\subsection{Deriving a local model for $\epsilon$ with enhanced dissipation}
\label{Sec_newdiss}

For the sake of physical completeness it would be preferable 
to switch to Eq.~(\ref{eq_epsilon}) and give up the local model 
Eq.~(\ref{eq_MLT_epsilon})--(\ref{eq_MLT_alpha}) altogether. However, as a first step into
that direction we can aim at modifying the computation of $\Lambda$ for the stably stratified layers
by guiding the necessary physical input through Eq.~(\ref{eq_epsilon}) and 
in particular through its local approximation,
Eq.~(\ref{eq_eps_alpha_NBV}). 
In a local framework we cannot accurately account for $D_f(\epsilon)$. Hence, we first
express $\tau$ in terms of $\Lambda$ in the local limit,
\begin{equation}
  \epsilon = \frac{2 K}{\tau} = c_{\epsilon} \frac{K^{3/2}}{\Lambda},
\end{equation}
from which we obtain that
\begin{equation}   \label{eq_tau_Lambda}
   \tau =  \frac{2}{c_{\epsilon}} \frac{\Lambda}{K^{1/2}}.
\end{equation}
To proceed we can now rewrite $c_3 \epsilon \tilde{N}$ as follows:
\begin{equation}   \label{eq_c3}
   c_3 \epsilon \tilde{N} = c_3 \frac{\epsilon}{\tau_{b}} = 2\,c_3 \frac{K}{\tau\,\tau_{\rm b}}.
\end{equation}
Following the analysis in the previous subsection we now compare Eq.~(\ref{eq_c3}) with 
\begin{equation}   \label{eq_c2}
   -c_2 \frac{\epsilon^2}{K} = -2\,c_2 \frac{\epsilon}{\tau}.
\end{equation}
In the stationary, local limit and assuming that we can absorb the contribution
from $\alpha_{\epsilon} \epsilon / \tau + 2\, c_1 g \alpha J / \tau$ into $-2\,c_2 \epsilon /\tau$
for sufficiently small $J$ and $\overline{w\epsilon}$ we obtain from 
Eqs.~(\ref{eq_eps_alpha_NBV}), (\ref{eq_c3}), and (\ref{eq_c2}) that
\begin{equation}   \label{eq_c3_c2_scale}
    \frac{c_3/\tau_{\rm b}}{2\,c_2/\tau} = \frac{c_3}{2\,c_2}\frac{\tau}{\tau_{\rm b}} \approx 0.078125 \frac{\tau}{\tau_{\rm b}} = \frac{25}{320} \frac{\tau}{\tau_{\rm b}} \approx 1,
\end{equation}
where the numerical value is obtained from setting $c_2=1.92$ and $c_3=0.3$. 
Contributions absorbed into the $-2\,c_2 \epsilon / \tau$ term could be accounted for
by small change of $c_2$. As inspection of the full Reynolds stress models solved in
\citet{kupka02b} demonstrates this is well justified since the two terms compared in
Eq.~(\ref{eq_c3_c2_scale}) completely dominate where $J < 0$.

This motivates the idea to also scale $\Lambda$, which according to Eq.~(\ref{eq_tau_Lambda})
is proportional to $\tau$, by a contribution $\propto \frac{25}{320} \frac{\tau}{\tau_{\rm b}}$. 
In GARSTEC the mixing length required for the turbulent convection model 
of \citet{kuhfuss1987} is computed following the prescription of \cite{wuchterl1995},
\begin{equation}   \label{eq_lambda_GARSTEC}
   \frac{1}{\Lambda} = \frac{1}{\alpha\,H_p} + \frac{1}{\beta_{\rm s}\,r},
\end{equation}
where $\beta_{\rm s}$ is a factor chosen to be $1$ in convectively unstable layers, where 
$\beta = -(dT/dr-(dT/dr)_{\rm ad}) > 0$ and thus $\nabla-\nabla_{\rm ad} > 0$, 
and $\beta_{\rm s}$ is possibly less than $1$ 
elsewhere. We now account for the effect of enhanced dissipation by gravity waves through reducing 
$\beta_{\rm s}$ to values less than 1. 
To this end we can interpolate between the two asymptotic cases 
$\tilde{N} \rightarrow 0$ and $\tilde{N} = \tau_{\rm b}^{-1} \gg \tau^{-1}$ through 

\begin{equation}   \label{eq_scale}
  \beta_{\rm s} = (1 + \lambda_{\rm s}\,\tilde{N})^{-1} \quad \mbox{\rm for} \quad M_r > M_{\rm schw}
\end{equation}
where $M_{\rm schw}$ is the mass of the convectively unstable core and thus identifies the mass shell
for which $\nabla = \nabla_{\rm ad}$ and $\lambda_s$ is a model parameter. Comparisons with solutions of the non-local Reynolds
stress model of \citet{canuto98b} for A-type stars \citep{kupka02b} show that $\tau_{\rm b} \approx 0.1\,\tau$ 
where $F_{\rm conv}$ reaches its negative minimum. This range of values for $\tau_{\rm b}$ is what we also
expect from Eq.~(\ref{eq_c3_c2_scale}) for a moderate variation of $c_2$.

\begin{figure*}
   \centering
   \includegraphics[width=0.90\textwidth]{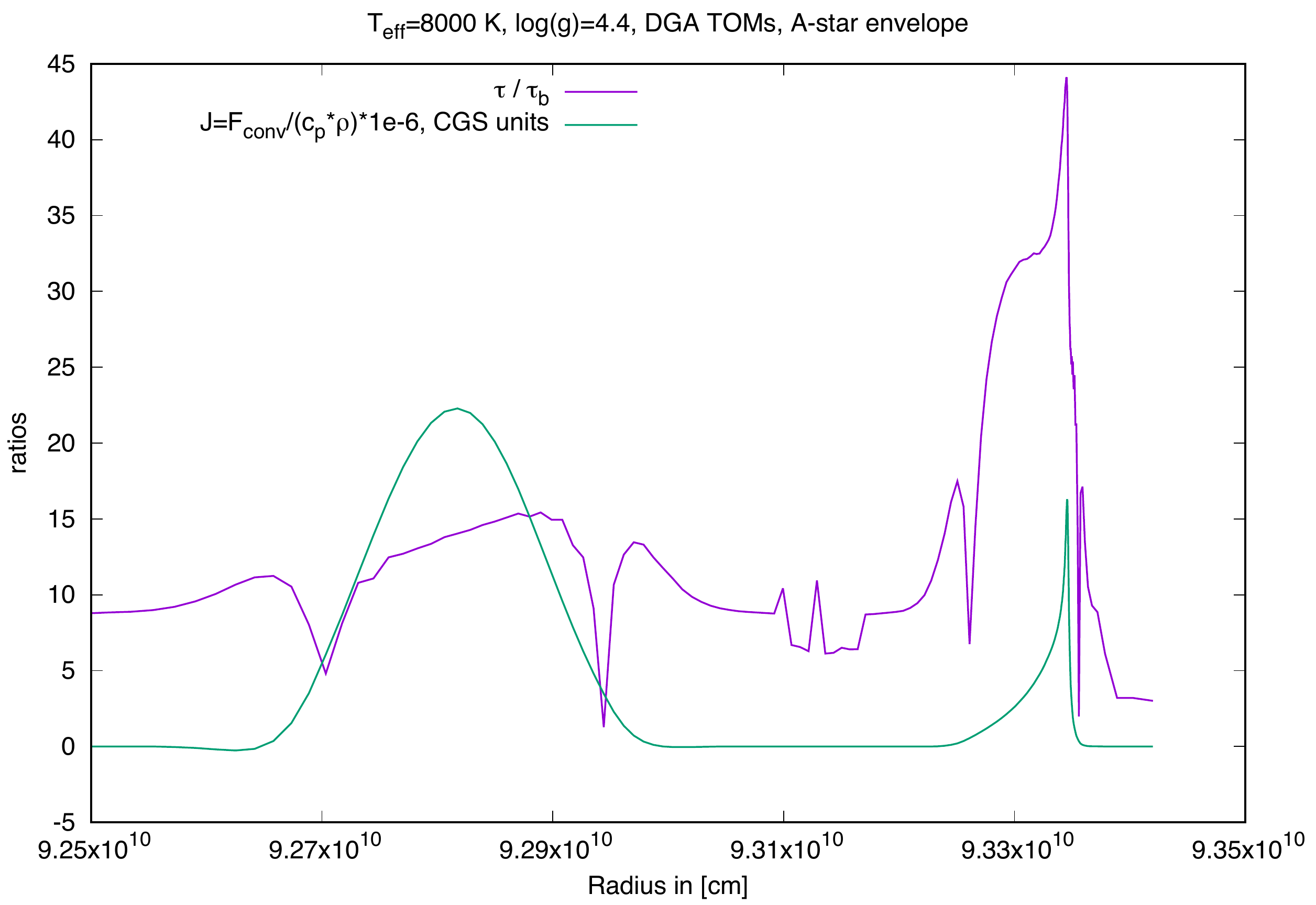}
      \caption{Ratio of $\tau/\tau_{\rm b}$ as a function of convective stability from a solution of
                   the non-local Reynolds stress model as presented in \citet{kupka02b} assuming the 
                   downgradient approximation for third order moments. The time scale $\tau_{\rm b}$ is computed
                   from $\tilde{N}^{-1}$ where the absolute value of $\beta$ is taken. Sign changes are hence
                   indicated by spikes. Both the overshooting zones below and above the lower and the upper
                   convectively unstable zone show the same increase of $\tau/\tau_{\rm b}$ from 0 to more than 10
                   (the finite grid resolution prevents $\tau/\tau_{\rm b}$ from becoming actually zero).
              }
         \label{fig2}
\end{figure*}

The results of \citet{kupka02b} can hence provide a rough guideline for the choice of $\lambda_{\rm s}$
and imply that $\Lambda$ is rapidly reduced by an order of magnitude {\em already within the countergradient region}
from the value it has at the Schwarzschild stability boundary (see Fig.~\ref{fig2}). This value is then maintained 
throughout the remainder of the countergradient region and the entire region where $F_{\rm conv} < 0$, in agreement 
with the $\tau\,\tilde{N} = O(1)$ suggested by \citet{canuto11d} in his Eq.~(5h). The preceding arguments and 
the analysis in the previous subsection show how this relation is connected with the full Eq.~(\ref{eq_epsilon})
and how this result can be implemented into a physically motivated reduction factor for
the mixing length through Eq.~(\ref{eq_lambda_GARSTEC}) and~(\ref{eq_scale}). Since the rough constancy 
of $\tau/\tau_{\rm b}$ (or the ``dominance'' of the term $c_3 \epsilon \tilde{N}$ in Eq.~(\ref{eq_epsilon}))
also causes the linear decay of the root mean square velocity as a function of distance in
the results of \citet{kupka02b} and \citet{montgomery04b}, and because the latter has also been recovered
from 3D radiation hydrodynamical simulations \citep{kupka18b} for just those layers, the entire procedure
is at least indirectly supported by this physically much more complete modelling. Similar results are not
yet available for convective cores, however.

In spite of its simplicity the disadvantage of Eq.~(\ref{eq_scale}) is the fact
that $\lambda_{\rm s}$ is a {\em dimensional} parameter. It hence has to be determined separately
for each stellar evolution model by numerical experiments which yield the value it has to have 
for a sufficient reduction of $\Lambda$ by an order of magnitude. For stars of different
mass this may have to be changed, and for later stages of stellar
evolution it will be even less convenient.
What we need here is an estimate for $\tau$. Without solving Eq.~(\ref{eq_epsilon}) this is akin
to a hen and egg problem, since in the end this would require just the quantity $\Lambda$ we are
up to compute: $\lambda_{\rm s} = (25/320) \,\tau$ with $\tau$ computed from Eq.~(\ref{eq_tau_Lambda}).
We could simplify this by setting $\tau = (2/c_{\epsilon}) (\alpha H_p K^{-1/2})$ or $\tau = (2/c_{\epsilon}) (r K^{-1/2})$,
as this formula is to be used only for $r>0$ and $H_p < \infty$ anyway. However, this has the
disadvantage that near the outer edge of the overshooting zone where $K \rightarrow 0$ one
obtains $\tau \rightarrow \infty$. From standard calculus applied to Eq.~(\ref{eq_lambda_GARSTEC})
we then obtain that $\Lambda \approx \alpha H_p$ right there which is exactly {\em not} what
we want. 
But we can rewrite Eq.~(\ref{eq_scale}) into 
\begin{equation}   \label{eq_scale2}
     \beta_{\rm s} = (1 + c_4 \Lambda K^{-1/2} \tilde{N})^{-1} \quad \mbox{\rm for} \quad M_r > M_{\rm schw}
\end{equation}
with
\begin{equation}  \label{eq_c4}
    c_4 = \frac{c_3}{2\,c_2} \frac{2}{c_\epsilon} \approx
    \frac{25}{320} \frac{2}{c_\epsilon} \approx \frac{5}{32 c_\epsilon} =
    0.19659 \approx 0.2,
\end{equation}
for which we have used $c_{\epsilon} = \pi (2/(3\,\rm Ko))^{3/2} \approx 0.7948 \approx 0.8$ with $\rm Ko=5/3$
from \citet{canuto98b}\footnote{If we used the value of $c_{\epsilon} \approx 2.18$ suggested in \citet{kuhfuss1987} we would instead obtain that 
		$c_4 \approx 0.07$. However, in the product $c_\epsilon K^{3/2}/\Lambda$ the constant $c_{\epsilon}$ to some extent cancels out, hence, the 
		overshooting distance is only weakly depending on this parameter. We discuss this further in Appendix~B of Paper~{\sc II}.}. 
This is achieved by realising that $\lambda_{\rm s}\,\tilde{N} = c_4 \Lambda K^{-1/2} \,\tilde{N}
= ((2 c_3) / (2 c_2 c_{\epsilon}) \Lambda K^{-1/2} \,\tilde{N} = (c_3 / (2 c_2)) \tau_{\rm b}^{-1}  (2/c_{\epsilon}) \Lambda K^{-1/2}
= ((c_3/\tau_{\rm b})/(2 c_2 / \tau))\cdot (2 \Lambda K^{-1/2} / (\tau c_{\epsilon})) = (c_3 / (2 c_2)) \cdot (\tau / \tau_{\rm b}))$ 
which is just Eq.~(\ref{eq_c3_c2_scale}) and where we have used Eq.~(\ref{eq_tau_Lambda}) for the last step.
Eq.~(\ref{eq_scale2}) is equivalent to Eq.~(\ref{eq_scale}) and also interpolates between the two asymptotic
cases, the transition between locally stable to unstable stratification ($\tilde{N} \rightarrow 0$) as well as
the overshooting region far away from the convective zone, where flow motions are dominated by waves 
($\tilde{N} = \tau_{\rm b}^{-1} \gg \tau^{-1}$).
Eq.~(\ref{eq_lambda_GARSTEC}) combined with Eq.~(\ref{eq_scale2})--(\ref{eq_c4})
can be rewritten into a quadratic equation for $\Lambda$ for which the positive branch can be taken
or which can be solved implicitly, for instance, by an iterative scheme (the former will be done in Paper~{\sc II}). 
In principle, the parameter $c_4$ could be adjusted to achieve the goal of $\tau_{\rm b} \approx 0.1\,\tau$ 
or rather $\Lambda(\min(F_{\rm conv})) \approx 0.1 \Lambda(M_r=M_{\rm schw})$ which mimics the result
discussed in Fig.~\ref{fig2} and in the previous paragraphs. However, we prefer to assume sufficient
generality of Eq.~(\ref{eq_epsilon}) and its parameters and therefore use them without further adjustments.
Some numerical experiments on the effects of varying $c_4$ can be found in Appendix~B of  Paper~{\sc II}.
In the next section we show that this procedure also leads to a finite overshooting layer 
which does not (notably) grow during stellar evolution.
\section{Discussion: Kuhfu{\ss} 3-equation model with enhanced dissipation}   \label{sect_discussion}
\subsection{Flow anisotropy instead of enhanced dissipation}  \label{sect_flowaniso}

A very important difference between the \cite{kuhfuss1987} and the
\cite{canuto98b} model is the set of convective variables
considered. In addition to the TKE \cite{canuto98b} also solve for
the vertical TKE. This means that the ratio of $\overline{w^2}/K$ is
not fixed a priori but is an outcome of the theory. \cite{kuhfuss1987}
on the other hand assumes full isotropy in the whole convection zone
which translates to a fixed ratio of
$\overline{w^2}/K=2/3$. Furthermore, the Kuhfu\ss~model uses an
isotropic estimate of the radial velocity
$v_\text{radial}=\sqrt{2/3\omega}$ in the non-local terms. Hence,
these terms are potentially overestimated by overestimating the ratio
of vertical to total kinetic energy. This could result in an
unreasonably large overshooting zone. The treatment of the flow
anisotropy is especially problematic at convective boundaries where
the flow turns over. In the convective boundary layers the motions
change from being predominantly radial to becoming predominantly
horizontal. This means that the ratio of vertical to total kinetic
energy should drop from the isotropic value to smaller
values.

\begin{figure}
	\centering
	\includegraphics{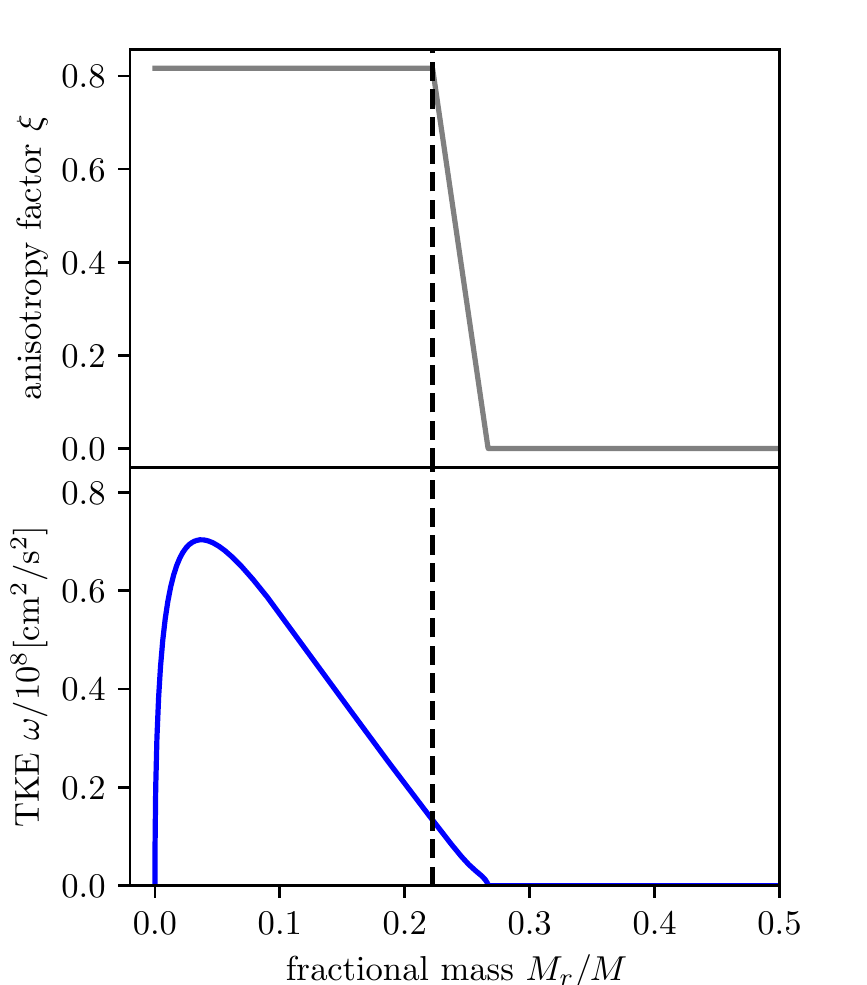}
	\caption{Artificial anisotropy factor $\xi$ and TKE as a function of fractional mass in the upper and lower panel, respectively. The black dashed line indicates the Schwarzschild boundary.}
	\label{figadhocanis}
\end{figure}

To study the impact of anisotropy we mimic the change of the flow
pattern by introducing an artificial anisotropy factor
$\xi^2=\overline{w^2}/K$. This anisotropy factor is set to a value of
$\xi=\sqrt{2/3}$ in the bulk of the convection zone and then linearly
decreases to a value of zero from the Schwarzschild boundary
outwards. This is most probably not a very physical behaviour but just
meant for illustrative purposes. The profile of this artificial
anisotropy factor is shown in the upper panel of
Fig.~\ref{figadhocanis}. The profile of the TKE computed with this
anisotropy factor is shown in the lower panel of the same figure. The
black dashed line indicates the Schwarzschild boundary. It can be seen
that an overshooting zone beyond the Schwarzschild boundary emerges,
which has, however, a clearly limited extent. As intended, a limitation of the 
anisotropy could solve the problems observed with the original version of the 3-equation model. 
The description requires another free parameter which is the slope of the linear function. 
The slope parameter directly controls the overshooting distance which is very similar to other 
ad hoc descriptions of convective overshooting. Also, the functional form of $\xi$ has not been 
determined by physical arguments but has been chosen arbitrarily.

This unfavourable situation should be avoided by a physically motivated estimate for the anisotropy factor. This requires to compute the vertical kinetic energy. To obtain an estimate of the distribution of the turbulent kinetic energy in the \cite{kuhfuss1987} model we start from the fourth equation of the \cite{canuto98b} model:
\begin{align}
\dpar{}{t}\frac{1}{2}\overline{w^2}+D_f\left(\frac{1}{2}\overline{w^2}\right)=-\frac{1}{\tau_{pv}}\left(\overline{w^2}-\frac{2}{3}K\right)+\frac{1}{3}(1+2\beta_5)g\alpha J-\frac{1}{3}\epsilon\label{eqvertkin}
\end{align}
which solves for the vertical turbulent kinetic energy
$\overline{w^2}$. Not solving for $\overline{w^2}$ implies that also $D_f\left(\frac{1}{2}\overline{w^2}\right)$ is unknown. A reasonable way to compute this quantity from the \cite{kuhfuss1987} model is again to assume an isotropic distribution of the fluxes: $D_f\left(\frac{1}{2}\overline{w^2}\right)=\frac{1}{3}D_f(K)$. By rearranging and neglecting the time-dependence in Eq.~(\ref{eqvertkin}) we can define an anisotropy factor:
\begin{align}
\frac{\overline{w^2}}{K}=\frac{2}{3}-\frac{\tau_{pv}}{K}\left(\frac{1}{3}D_f(K)-\frac{1}{3}(1+2\beta_5)g\alpha J+\frac{1}{3}\epsilon\right)\label{eqanis}
\end{align}
All quantities in Eq.~(\ref{eqanis}) can be computed within the Kuhfu\ss~3-equation model.

\begin{figure}
	\centering
	\includegraphics{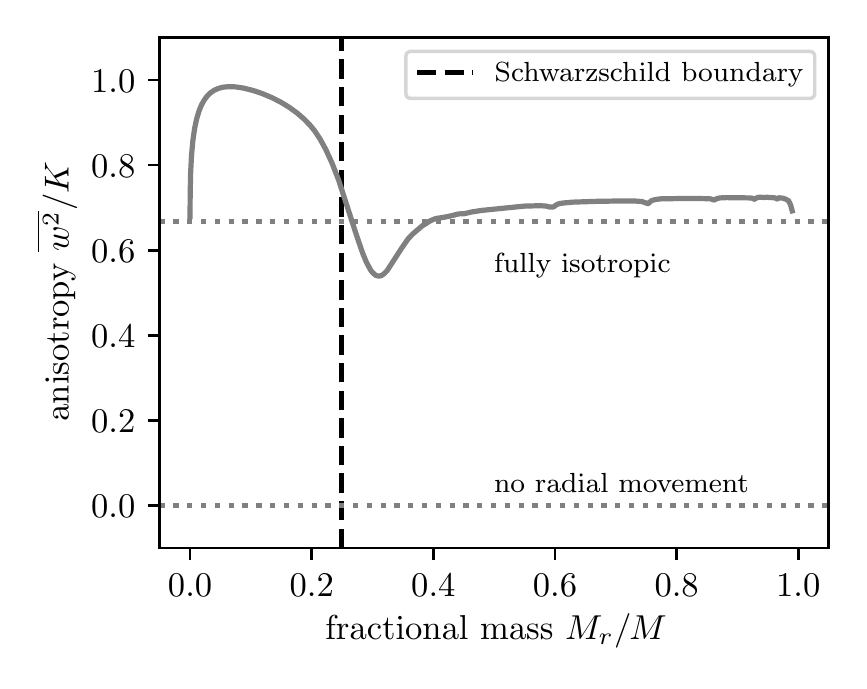}
	\caption{Estimate of the anisotropy factor according to Eq.~(\ref{eqanis}) for a 3-equation model without limited dissipation length-scale $\Lambda$. 
	The profile of the turbulent kinetic energy of this model is shown in Fig.~\ref{figenergyorig}.
	}
	\label{figaniso}
\end{figure}

We have computed the anisotropy factor according to Eq.~(\ref{eqanis}) for a stellar model which used the original version 
of the Kuhfu\ss~3-equation model. The result is shown in Fig.~\ref{figaniso}. In the bulk of the convection zone within the Schwarzschild boundary the estimated anisotropy 
points towards a radially dominated flow. Directly beyond the Schwarzschild boundary the estimated anisotropy factor drops below 
the isotropic value of 2/3. This can be attributed to the negative convective flux in the overshooting zone which according to 
Eq.~(\ref{eqanis}) reduces the ratio of vertical to total kinetic energy. Further out in mass coordinate the estimated anisotropy 
increases again slightly above a value of 2/3 and remains to a good approximation constant over the region in which positive 
kinetic energy is observed (see Fig.~\ref{figenergyorig}).

Introducing this anisotropy factor into the Kuhfu\ss~3-equation model would not substantially reduce the estimate of the radial velocity. 
On the contrary, over large parts of the model the value of the radial velocity would be even larger than the current estimate as 
we find an anisotropy factor above the isotropic value of 2/3. To finally settle the question of the flow anisotropy in Reynolds stress models 
one also has to solve the respective equation for the vertical kinetic energy (Eq.~(\ref{eqvertkin}) shown here, as taken
from the \citealt{canuto98b} model) self-consistently 
coupled to the non-local convection model. 
However, since such a more realistic anisotropy factor cannot resolve the problem of 
excessive mixing found in the original Kuhfu\ss~3-equation model and because its implementation
as an additional differential equation increases the complexity of the model, we first perform
a thorough analysis of the improved 3-equation model in Paper~{\sc II} and postpone the extension
of this new model to future work.

\subsection{Dissipation in the Kuhfu\ss~1- and 3-equation model}

We have implemented the enhanced dissipation mechanism, developed in Sect.~\ref{Sec_newdiss}, into GARSTEC. For the details of the implementation we here refer to Paper~II. With this implementation we solve the stellar structure equations and the convective equations (\ref{eqKuh1}) - (\ref{eqKuh3}) self-consistently. 
{We note that for consistency and to simplify the comparison between
the 1-equation and the 3-equation model, we set $c_{\epsilon}=C_D$
(see Appendix~\ref{secKuh}), whence it follows that $c_4 \approx 0.072$ in those
calculations.
As an example we show here the TKE in a $5\,M_\odot$ main-sequence star in Fig.~\ref{figenergyextended}. The Schwarzschild boundary is indicated with a black dashed line. In this model the convective energy extends slightly beyond the Schwarzschild boundary which means that an overshooting zone emerges consistently from the solution of the model equations. However, in contrast to Fig.~\ref{figenergyorig} the energy does no longer extend throughout the whole star but has a clearly limited extent as one would expect for this kind of star in this evolutionary phase.

This shows already that the enhanced dissipation mechanism proposed
above is able to solve the problems observed in the original version
of the 3-equation Kuhfu\ss~convection model. The detailed structure
and the behaviour of stellar models with different initial masses will
be discussed in Paper~II.
\begin{figure}
	\centering
	\includegraphics{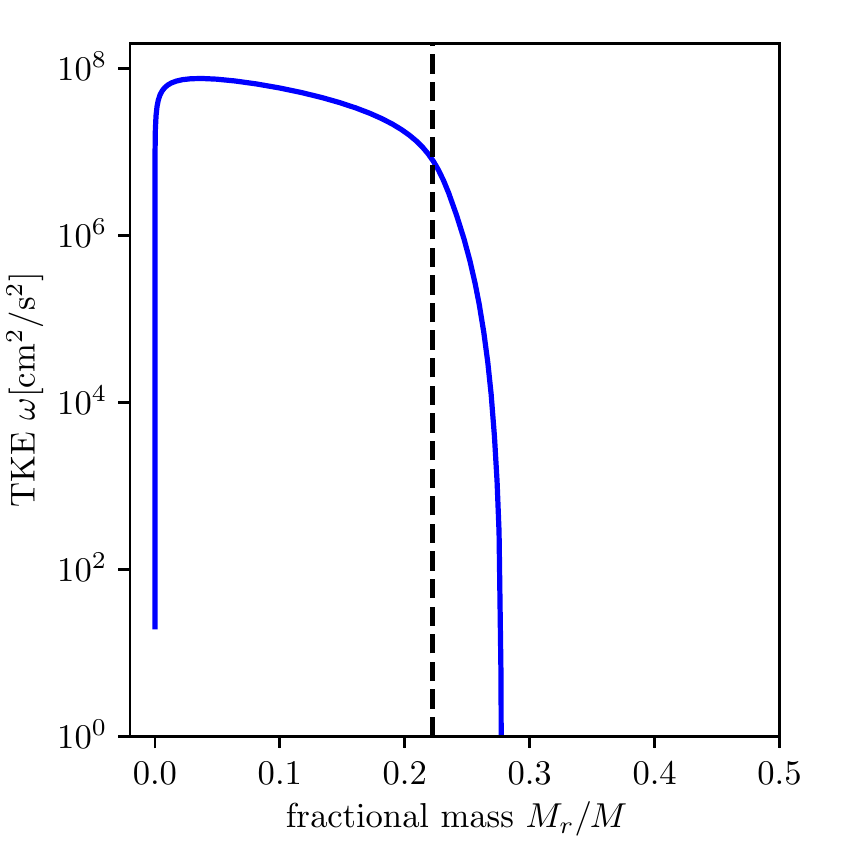}
	\caption{Convective energy as a function of the fractional mass for the Kuhfu\ss~model 
	             including the improved dissipation mechanism. The Schwarzschild boundary is 
	             indicated by a dashed black line.}
	\label{figenergyextended}
\end{figure}

The results obtained from the different versions of the Kuhfu\ss~model
can be interpreted by studying the individual terms of the TKE
equation (Eq.~\ref{eqKuh1}) in more detail. In
Fig.~\ref{figtermcompare} we show the three terms of the TKE
equation---buoyant driving, dissipation and non-local flux---with a corresponding
red, black, and blue line respectively for the 1-equation model (panel
a), the original 3-equation model (panel b) and the improved
3-equation model (panel c).

Stellar models applying the non-local 1-equation theory posses a clearly bounded convective region with a reasonable extent.
However, this is achieved by suppressing the countergradient layer and artificially coupling the sign of
the convective flux to that one of the superadiabatic gradient.

When using the 3-equation model in its original version this welcome property vanishes and the stellar models become fully convective. 
As discussed in Appendix~\ref{secKuh} the 3-equation model does not approximate the convective flux by a local model but rather solves 
an additional differential equation for it. This reduces the coupling of the different convective variables. Intuitively one would expect this 
model to be physically more complete than the 1-equation model and to yield physically improved models (see the discussion in Sect.~5 of \citealt{kupka20b}). 
However, the stellar models computed with the 3-equation model look physically unreasonable,
as the existence of fully convective B-stars with $5\,M_{\odot}$ is
excluded from the lack of stars hotter than the hydrogen main-sequence. 

This rises the question why a seemingly physically more complete model leads to worse results. 
It can be illustrated by comparing the TKE terms in the 1- and original 3-equation models shown in panels a) and b) in Fig.~\ref{figtermcompare}. 
In the 1-equation model the buoyant driving term which is proportional to the convective flux shows negative values in the overshooting zone, 
which is expected due to the buoyant braking in the stable layers. The buoyant term even exceeds the actual dissipation term in magnitude. 
This means that in the 1-equation model it is  not the dissipation term but rather the buoyant driving term which acts as the main sink term 
in the overshooting zone. When applying the 3-equation model the buoyant term is still negative in the overshooting zone. The values are, however, 
much smaller in magnitude compared to the 1-equation model. The dissipation and non-local flux term have about the same magnitude 
in the overshooting zone as obtained with the 1-equation model, because their functional form did not change. Considering that it was 
the buoyant driving term which was acting as the main sink term, the 3-equation model in its original form is lacking a sink term in the overshooting zone. 
This naturally explains the excessive overshooting distance found for this model.

To understand how the dissipation by buoyancy waves can mitigate this problem it is worth to recall the approximation for the convective flux in the 1-equation model. \cite{kuhfuss1987} has approximated this to be $\Pi\propto(\nabla-\nabla_\text{ad})$. As the convective flux is the major sink term in the overshooting zone in the 1-equation model one possibility is to introduce a dissipation term which has the same dependence, $\epsilon\propto(\nabla-\nabla_\text{ad})$. A process with this dependence would be, for example, the dissipation by buoyancy waves as proposed above. We have demonstrated that the enhanced dissipation by buoyancy waves reduces the overshooting distance again to a more reasonable extent for the TKE (see Fig.~\ref{figenergyextended}). The related terms of the TKE equation are shown in Fig.~\ref{figtermcompare} in panel c). In the overshooting zone the magnitude of the dissipation term is now substantially larger than the negative buoyancy term such that it acts as the dominant sink term. Also the shape of the dissipation profile has changed compared to the original 3-equation case. The transition from finite to zero values looks smoother for the improved 3-equation model because the temperature gradient which has readjusted differs in comparison with the 1-equation model. 

This comparison shows why the original version of the 3-equation model results in fully convective stars. The fact that a sink term is missing 
points again at the importance of a dissipation term which is proportional to $(\nabla-\nabla_\text{ad})$. On a first glance, a negative convective flux 
with larger magnitude in the overshooting zone could also increase the sink term in the TKE equation. But the following line of arguments shows that this 
hypothesis leads to unplausibly large non-local fluxes. 
\begin{figure}
	\centering
	\includegraphics{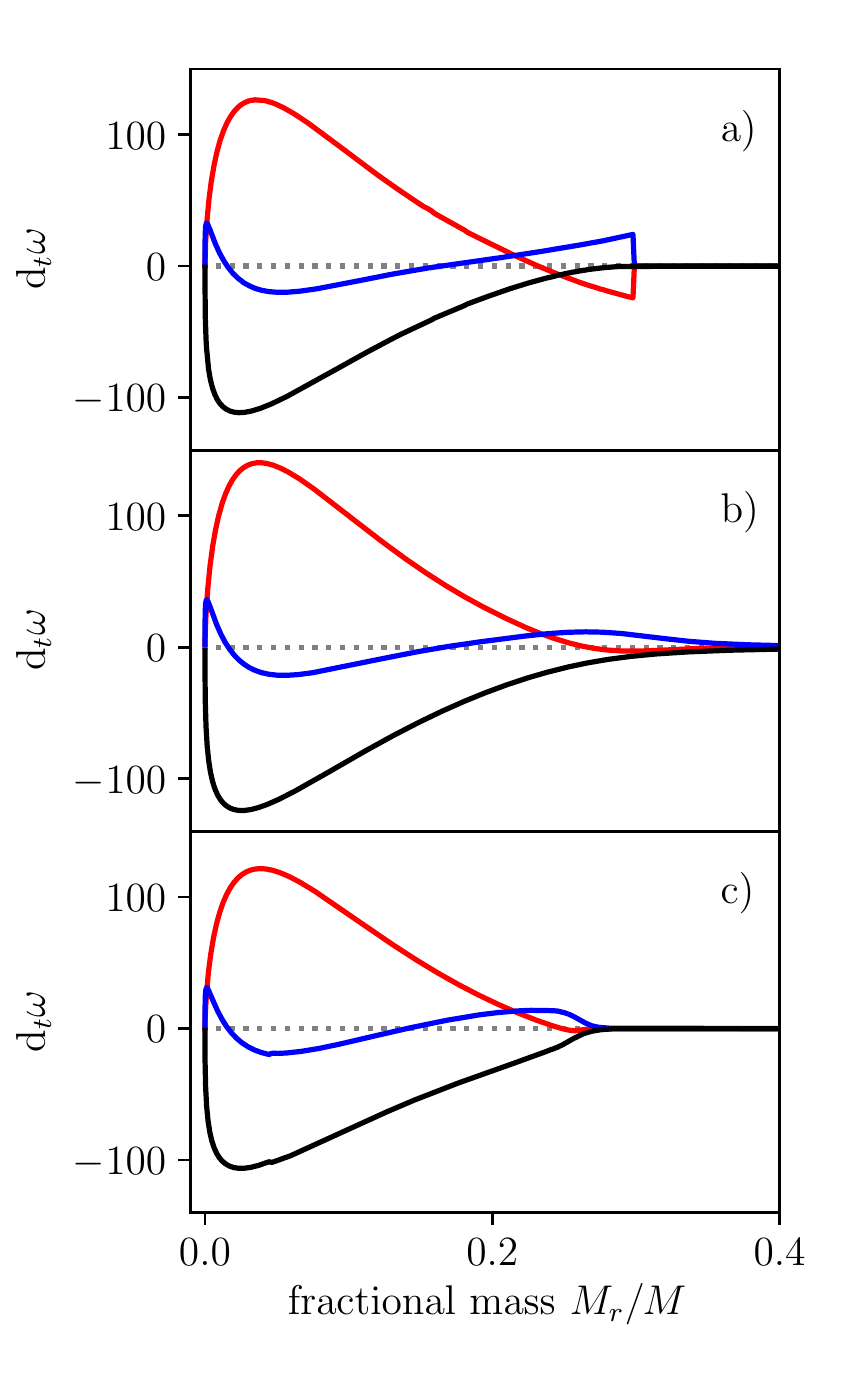}
	\caption{Comparison of the different terms in the TKE equation (Eq.~\ref{eqKuh1}) in the Kuhfu\ss~ 1-equation (panel a), original 3-equation (panel b) and improved 3-equation (panel c) model. The buoyant driving term, the dissipation term and the non-local flux term are shown with a red, black, and blue line here.}
	\label{figtermcompare}
\end{figure}

Here, we consider Eq.~(\ref{eqKuh1})--(\ref{eqKuh3}).\footnote{We point out that exactly the same sequence of arguments applies to the equivalent three 
       equations for the turbulent kinetic energy $K=\overline{q^2}/2$, the squared fluctuation of the difference between local temperature and its Reynolds 
       average, $\overline{\theta^2}$, and the cross correlation between velocity and temperature fluctuations, $J=\overline{w \theta}$, as they appear in the model 
       of \citet{canuto98b} and discussed in \citet{kupka20b}.}
Let us assume that $J$ becomes larger, or, equivalently, $\Pi$ in Eq.~(\ref{eqKuh1})--(\ref{eqKuh3}) becomes larger in magnitude in the region 
where it is negative. Then, the buoyant driving term shown in panel~b of Fig.~\ref{figtermcompare} changes towards more negative values. This permits the source, 
the divergence of the flux of kinetic energy, to become smaller. However, in that case the buoyant driving term (containing $\Pi$) also becomes larger in 
Eq.~(\ref{eqKuh3}), which predicts the magnitude of entropy fluctuations.

Since the vertical velocities have to become smaller, when the non-local flux of kinetic energy becomes smaller (and we assume a constant 
anisotropy in this thought experiment), the squared fluctuations of entropy, $\Phi$, or of temperature, $\overline{\theta^2}$, have to become {\em larger} instead.
But for $\Pi < 0$ in the region we consider here, both $-\Pi/\tau_{\rm rad}$ and $(2\nabla_\mathrm{ad}T / H_p) \Phi$ act as sources which are boosted in Eq.~(\ref{eqKuh2}).
Unless we would consider a large rate of change in the non-local transport of convective flux and entropy fluctuations, the only way to obtain
an equilibrium solution in this model is to increase velocities and thus also the flux of kinetic energy. This is exactly the solution observed in
panel~b of Fig.~\ref{figtermcompare} with its excessively extended overshooting. The closure used in \citet{canuto93b} and \citet{canuto98b}, which also accounts 
for buoyancy contributions to the correlation between fluctuations of temperature and the pressure gradient (the $-\Pi/\tau_{\rm rad}$ term in Eq.~(\ref{eqKuh2})) 
does not change this argument. But a scenario that builds up large fluctuations of entropy in the overshooting region, where radiative cooling should efficiently 
smooth them while it has to suppress high velocities, appears
unphysical. Thus, this alternative can be excluded.

Since extensive overshooting, which eventually mixes the entire B-star, is ruled out by observations, we are left with flow anisotropy or enhanced dissipation 
due to the generation of waves as physical mechanisms to limit overshooting in the 3-equation framework. Because extreme levels of flow anisotropy are neither found
in solar observations nor in numerical simulations of overshooting in white dwarfs \citep{kupka18b}, nor in solutions of the model of \citet{canuto98b} 
for A-stars \citep{kupka02b} or white dwarfs \citep{montgomery04b}, there is hardly evidence for this idea. On the contrary, the enhanced energy dissipation rate is 
contained in the full model of \citet{canuto98b} which yields at least some qualitative agreement with numerical simulations of several scenarios of stellar overshooting
(see \citealt{kupka02b} and \citealt{montgomery04b} and compare with \citealt{kupka18b} for the latter). This makes the improved computation of the dissipation rate 
of kinetic energy the most plausible improvement of the 3-equation model to remove the deficiency the model has had in its original version proposed by \citet{kuhfuss1987}.

\subsection{Comparing the Kuhfu\ss~3-equation model with overshooting models and numerical simulations}   \label{sect_discussion_modsim}

\citet{viallet2015} have reviewed several models suitable for parametrization of 
overshooting above stellar convective cores. One of them is the model proposed by 
\citet{freytag96b} based on 2D hydrodynamical simulations of thin convective zones 
which appear in the atmosphere and upper envelope of stars. The simulations had to be restricted 
to low Peclet (Pe) numbers where highly efficient radiative diffusion competes with convective 
energy transport. The simple exponential decay law for velocity as a function of distance from 
the convection zone has been particularly attractive for stellar evolution modelling and 
the model is available in most actively used stellar evolution codes including GARSTEC. 
As the velocity scales with the pressure scale height, this model requires an additional 
cut-off to prevent diverging overshoot from very small convective cores as found in
stars with less than two solar masses. We will discuss this issue in detail in Paper~{\sc II}. 
Additionally, \citet{kupka18b} have pointed out that within the countergradient and plume 
dominated regions of convective overshooting zones exponential decay rates for velocity 
work only within a limited spatial range \citep[see also][]{montgomery04b}.  
\citet{cunningham19b} argued for different decay rates for the plume dominated
and the wave dominated regime. Such distinctions are, however, not made in 
applications of that model. We refer to Paper~{\sc II} to a detailed comparsion 
of convective core sizes between the Kuhfu\ss~3-equation model and the exponential 
overshooting model, and here just emphasize that the energy loss of turbulent flows 
due to waves is readily built into the improved Kuhfu\ss~3-equation model.

Another model, suitable for a higher Pe regime, where penetrative
convection due to plumes occurs, is the one originally suggested by \citet{zahn91b}. 
When applied to convective cores his model had to rely on invoking Roxburgh's integral 
constraint \citep{roxburgh89b} for self-consistent predictions which effectively turns it 
into a model similar in complexity to the 1-equation model by \citet{kuhfuss1987}. We recall 
here that the 3-equation model with enhanced dissipation has a built-in dependence on Pe 
by accounting for radiative losses in its dynamical equations. A detailed discussion 
on the role of Pe in the 1-equation and 3-equation models can also be found in 
Paper~{\sc II}. The latter model is also not subject to the simplifications made in 
\citet{roxburgh89b} concerning the treatment of the dissipation rate $\epsilon$.

Finally, for the very high Pe regime of convective entrainment \citet{viallet2015} 
considered a model based on estimates relying on the variation of the inverse buoyancy 
time scale in the stably stratified layer next to a convective zone and the kinetic energy 
available at the boundary of the convective zone. The Kuhfu\ss~3-equation model can 
also deal with this case since it is the regime in which heat conduction is negligibly small. 
Hence, instead of relying on physically different models which have not been designed to be 
compatible among each other, the new 3-equation model can deal with the different regimes 
discussed in \citet{viallet2015} within a single formalism and without the necessity of fine 
tuning for these different cases.

Comparing the predictions of the new model with those concluded from 3D hydrodynamical simulations 
of convection is more difficult: as already mentioned in Sect.~\ref{sect_reduced_ML} they currently have
to be restricted to a different parameter range. In \citet{kupka20b} it is explained why the effective (numerical)
heat conductivity in the simulations has to be higher than the physical one which leads to values of Pe several orders of
magnitudes smaller than those found in stars. As the numerical diffusion of momentum and heat in high Pe 
simulations have to remain comparable to each other, we have to expect differences in flow structures 
and overshooting distances when compared to the actual, stellar parameter range  (see, for instance, 
\citealt{scheel17b} and \citealt{kp19b}). Nevertheless, it is a very important finding for the veracity
of the enhanced Kuhfu\ss~3-equation model that the 3D simulation results concerning convective cores 
by \citet{browning04b}, \citet{gilet2013}, \citet{Rogers13b}, \citet{Augustson16b}, and \citet{edelmann2019},
among others, and the related simulations of convective shells by \citet{meakin2007}, all show that convective 
zones excited by nuclear burning are subject to convective entrainment and penetration, respectively, depending 
on the specific setup, and in each case gravity waves are excited which extend throughout the radiative stellar 
envelope. This supports the theoretical analysis of \citet{linden75b} and \citet{zeman77b} for the equivalent 
scenario in meteorology which lead to the non-local dissipation rate equation proposed in \citet{canuto94b} 
and generalized to applications in stellar convection by \citet{canuto98b}, see Eq.~(\ref{eq_epsilon}), which is 
the starting point for our investigations we detail in this paper.

\section{Conclusions}
\label{sect_conclusions}

The original model by \citet{kuhfuss1987} was shown by \citet{flaskamp2003} to lead to
convective overshooting zones on top of convective cores that fully 
mix the entire object on a fraction of its main sequence life time. We verified that
the ad hoc cure to reduce the ratio of vertical to total TKE to zero
no longer works once realistic models for that quantity are used. 
From a physical point of view the ad hoc cure is hence
ruled out as an explanation for this deficiency of the model by \citet{kuhfuss1987}.
In this paper a physically motivated modification of the mixing length 
has hence been suggested which takes into account that the dissipation rate
of TKE has been underestimated by the original 3-equation model of \citet{kuhfuss1987}. 
In Paper~{\sc ii} we present more detailed tests of the improved 3-equation model 
proposed in this paper based on stellar evolution tracks for A- and B-type main sequence
stars of different masses.

One conclusion from these analyses appears to be that the minimum physics
to obtain realistic models of overshooting layers require to account for non-locality
of the fluxes of kinetic energy and potential temperature (as intended by \citealt{kuhfuss1987})
and in addition to account for the variation of the anisotropy of turbulent kinetic energy
as a function of local stability and non-local transport. If the latter is done in a realistic
way, it becomes also clear that a physically more complete model of the dissipation
rate of TKE is needed. All these features are already provided by the
model of \citet{canuto98b} which in its most simple form accounts for non-locality
with the downgradient approximation (as in the model of \citealt{kuhfuss1987}). The
present simplification is an attempt to carry over the most important features of
the more complete model by \citet{canuto98b} into the \citet{kuhfuss1987} model
which is already coded within GARSTEC.

Switching to more complex non-local convection models in a stellar evolution
code is not an easy task. This requires that the model and its implementation also account 
for the following:
\begin{enumerate}
\item Realistic, mathematically self-consistent boundary conditions. This is taken
        care of in the current implementation of the \citet{kuhfuss1987} model in GARSTEC.
\item A fully implicit, relaxation based numerical solver for the resulting set of equations.
        This is fulfilled by GARSTEC as well. Adding further differential equations always
        means some non-trivial work on this side.
\item A stable, monotonic interpolation scheme for the equation of state.
        Again this is  fulfilled in GARSTEC \citep{weiss2008}.
        If this is not fulfilled, $\beta$ cannot be computed correctly and
        any closure depending on its sign becomes uncertain, since oscillations may be
        fed into its computation.
\item A robust formulation of the dynamical equations which avoids cancellation errors
        introduced through a nearly perfectly adiabatic stratification. This is 
        realised in the implementation of the \citet{kuhfuss1987} model in GARSTEC indicated by
        the smoothness of the equation terms in Fig.~\ref{figtermcompare}. This can be attributed
    	to the fact that the implementation uses Eq.~(\ref{eq3eqnabla}) to compute the temperature 
    	gradient instead of numerical derivatives.
\end{enumerate}

Naturally, as discussed in \citet{ireland18b}, in \citet{augustson19b}, and in \citet{korre21b},
among others, rotation and magnetic fields influence convection and convective overshooting.
A path towards including rotation in non-local convection models has been investigated, e.g., 
by \citet{canuto98c} and by \citet{canuto11b}, but such extensions have to be left
for future work: the present model is only a first step beyond MLT-like models.

If the modified mixing length Eq.~(\ref{eq_lambda_GARSTEC}) and~(\ref{eq_scale}) 
and even more so Eq.~(\ref{eq_lambda_GARSTEC}) with Eq.~(\ref{eq_scale2})--(\ref{eq_c4})
turns out to produce stable, physically meaningfully evolving overshooting zones
with GARSTEC, further tests of this approach are highly warranting. These may
also motivate the implementation of fully non-local Reynolds stress models
at the complexity level of \citet{canuto98b} which completely avoid the introduction
of a mixing length with all its shortcomings.

\begin{acknowledgements}
F.~Kupka is thankful to the Austrian Science Fund FWF for support through projects P29172-N and P33140-N and 
support from European Research Council (ERC) Synergy Grant WHOLESUN \#810218. F.~Ahlborn  
thanks Martin Flaskamp for his pioneering work on the 3-equation non-local model by R.~Kuhfu{\ss}.
\end{acknowledgements}
\bibliographystyle{bibtex/aa} 
\bibliography{hydro,lrr-stellar_convection-refs,lrr-stellar_convection-refs_part2}
\appendix
\section{The Kuhfu{\ss}\  convection model}  \label{secKuh} 

In this appendix we summarise the turbulent convection model developed by \cite{kuhfuss1987} who
derived dynamical equations for three of the second order moments to model turbulent convection 
in the stellar interior: the turbulent kinetic energy, the turbulent convective flux, and the squared entropy 
fluctuations. The (specific) turbulent kinetic energy (TKE) is denoted by $K$ in the main text. Here,
we summarise those equations as used inside GARSTEC. They model entropy fluctuations instead 
of temperature fluctuations. To avoid confusion with other models and their implementation 
here we stick to the notation of \cite{kuhfuss1987}: TKE 
is denoted by $\omega$. The radial component of the turbulent convective flux is written as 
$\Pi$ and is computed from entropy fluctuations, consistent with choosing the squared entropy 
fluctuations $\Phi$ as the third dynamical variable of the system. Hence, the Reynolds splitting 
is performed for
\begin{align*}
\vec v=\overline{\vec v}+\vprime, \,\,\,\, \rho=\overline{\rho}+\rho', \,\,\,\, s=\overline{s}+s', \,\,\dots
\end{align*}
and the second order moments are computed from
\begin{align*}
\omega=\overline{\vprime^2/2},\,\,\,\, \Pi=\overline{\vprime\cdot s'}_r,\,\,\,\, {\rm and}\,\,\,\, \Phi=\overline{s'^2/2}.
\end{align*}
As for any TCM a number of assumptions and approximations is required to obtain closed
systems of equations that can actually be applied in stellar structure and evolution models. 
In the following we will briefly review the key assumptions of the Kuhfu\ss\ model. 
By using only the total TKE $\omega$ the \cite{kuhfuss1987} model is not able to account for a variable distribution 
of the kinetic energy in radial and horizontal directions. Instead the distribution of kinetic energy in radial and horizontal directions is assumed 
to be isotropic at all radii, such that one third of the energy is attributed to each spatial direction. The Kuhfu\ss~ model further neglects turbulent 
pressure fluctuations. As pointed out by \cite{viallet2013} pressure fluctuations play an important role for convection in envelopes, hence the 
Kuhfu\ss~ model is probably not suited to model envelope convection. Finally \cite{kuhfuss1987} also made use of the Boussinesq approximation. 
In the current implementation suggested by \cite{flaskamp2003} we also neglect effects due to the chemical composition, e.g. composition gradients.
\subsection{Viscous dissipation}

In the Kuhfu{\ss}\ model most terms containing the molecular viscosity are neglected because they are of minor importance compared to competing terms. 
Only the viscous dissipation term for the kinetic energy is considered to be non-negligible. \cite{kuhfuss1987} models the dissipation of the kinetic energy 
with a Kolmogorov-type term \citep{kolmogorov1968,kolmogorov1962}:
\begin{align}
\epsilon=C_D\frac{\omega^{3/2}}{\Lambda},
\label{eqdiss}
\end{align}
where $C_D$ is a parameter. \cite{kuhfuss1987} suggests a value of $C_D = 8/3\cdot\sqrt{2/3}$ to be compatible
with MLT in the local limit of his model.

In the Kolmogorov picture kinetic energy is dissipated thanks to a cascade through which energy is transferred to smaller and smaller spatial scales. 
The rate at which this dissipation happens is dominated by the largest scales at which energy is fed into the cascade. In Eq.~(\ref{eqdiss}) 
the length-scale $\Lambda$ refers to this largest scale of the turbulent cascade. As in the mixing length theory the length-scale is parametrised 
using the pressure scale height $H_p$ and an adjustable parameter $\alpha$: $\Lambda=\alpha H_p$. Problems with this parametrisation 
are discussed in the main text.

\subsection{Radiative dissipation}

Convective elements lose energy through radiation. This is considered in the energy conservation equation by including radiative fluxes 
as sink terms. In the Kuhfu{\ss}\ equations the radiative losses finally appear as dissipation terms:
\begin{align*}
\epsilon_{\mathrm{rad},\Pi}=\frac{1}{\tau_\mathrm{rad}}\Pi\,,\,\,\,\,\,\epsilon_{\mathrm{rad},\Phi}=\frac{2}{\tau_\mathrm{rad}}\Phi,
\end{align*} 
where \cite{kuhfuss1987} models radiative dissipation by introduing the radiative time-scale $\tau_\text{rad}$, which he defines as:
\begin{align*}
\tau_\text{rad}=\frac{c_p\kappa\rho^2\Lambda^2}{4\sigma T^3\gamma_R^2}.
\end{align*}
Here, $\gamma_R$ is a parameter which \cite{kuhfuss1987} sets to $2\sqrt{3}$ , again to recover the MLT model in the local limit.
Furthermore, $c_p$ refers to the specific heat capacity at constant pressure, $\kappa$ to Rosseland opacity and $\sigma$ to the 
Stefan-Boltzmann-constant. The variables $T$ and $\rho$ are temperature and density, as usual in stellar structure models.

\subsection{Higher order moments}

The Navier-Stokes equations contain non-linear advection terms. When constructing the equations for the second order moments 
these advection terms give rise to third order moments (TOMs). These higher order moments are the source of the non-local behaviour 
of the convection model. They can be cast into the form:
\begin{align*}
\mathcal{F}_a & = \frac{1}{\overline{\rho}}\operatorname{div}(j_a) \,\,\,\, {\rm with} \,\,\,\, j_a  = \overline{\rho}\,\overline{\vprime a},
\end{align*} 
where $a$ is a second order quantity. The closure of these TOMs is one of the main challenges of any TCM. 
\cite{kuhfuss1987} closes the system of equations at second order and describes each TOM using the so-called down-gradient 
approximation \citep[e.g.,][]{daly1970,launder1975,xiong78b,li2007}.  In the down-gradient approximation the fluxes $j_a$ are modelled 
following Fick's law:
\begin{align}
 \vec j_a & = -D_a\nabla \overline{a}, \\\label{eqfluxomega}
 D_a & = \alpha_a\overline{\rho}\Lambda \sqrt{\omega}.
\end{align}
This approximation is applied for the TOMs appearing in the equations for $\omega$, $\Pi$, and $\Phi$ with $a=\vprime^2/2$, 
$\vprime s'$, or $s'^2/2$. The parameters $\alpha_a$ control the impact of the non-local terms. \cite{kuhfuss1987} suggests 
a default value of $\alpha_\omega\approx0.25$. The values for the parameters $\alpha_{\Pi,\Phi}$ are calibrated to MLT in a local version 
of the Kuhfu{\ss}\ theory. However, no values for the non-local case are provided.

Alternatively, one could compute the TOMs by deriving equations for them in the same way as for the second order moments. 
This has been shown in \cite{canuto92b,canuto93b}, \cite{canuto98b}, or \cite{xiong1997}, for example, and introduces fourth order moments 
which again have to be closed.

\subsection{Final model equations}

The above listed approximations are implemented in the derivation of the Kuhfu{\ss}~model. The final set of partial differential equations reads:
\begin{align}
 \text{d}_t\omega&=\frac{\nabla_\mathrm{ad}T}{H_p}\Pi-\frac{C_D}{\Lambda}\omega^{3/2}-\mathcal{F}_\omega,\label{eqKuh1}\\
 \text{d}_t\Pi&=\frac{2\nabla_\mathrm{ad}T}{H_p}\Phi+\frac{2c_p}{3H_p}(\nabla-\nabla_\mathrm{ad})\omega-\mathcal{F}_\Pi-\frac{1}{\tau_\text{rad}}\Pi,\label{eqKuh2}\\
 \text{d}_t\Phi&=\frac{c_p}{H_p}(\nabla-\nabla_\mathrm{ad})\Pi-\mathcal{F}_\Phi-\frac{2}{\tau_\text{rad}}\Phi,\label{eqKuh3}
\end{align} 
where $\nabla$ and $\nabla_\mathrm{ad}$ refer to the model and adiabatic temperature gradient, respectively. The substantial derivative 
is defined as $\mathrm{d}_t=\partial_t+\overline{\vec v}\cdot\nabla$. For more details about the derivation we refer to the original work by \cite{kuhfuss1987} 
and \cite{flaskamp2003}.

Using the convective flux from the convection model one can compute the temperature gradient of the stellar model self-consistently from
\begin{align}
\nabla=\nabla_\text{rad}-\frac{H_p\rho}{k_\text{rad}}\Pi,
\label{eq3eqnabla}
\end{align}
with 
\begin{align*}
    k_\text{rad}=\frac{4acT^3}{3\kappa\rho}\,\,.
\end{align*}
where $a$ and $c$ denote the radiation constant and the speed of light respectively. Here, we neglect the kinetic energy flux $\vec j_\omega$, which is assumed to be small compared to the convective flux. Equation~(\ref{eq3eqnabla}) 
couples the convection model to the stellar structure equations. The self-consistent computation of the temperature gradient allows to study its behaviour 
in the overshooting region. This is an advantage over ad hoc descriptions of overshooting in which the temperature gradient is set manually.

\section{Alternatives to improve Eq.~(\ref{eq_epsilon})} \label{secEq11}

Eq.~(\ref{eq_epsilon}) is heavily parametrised. \citet{canuto09b} hence discussed
a number of simplified models used in geophysics for the computation of $\epsilon$.
They are based on modified mixing lengths which account for physical processes relevant
to dissipation. However, those models are not directly applicable to stellar convection: 
some of them consider a solid wall as a boundary and none of them has been designed 
for the extreme density contrast of deep stellar envelopes or the peculiarities of convective 
cores in massive stars. 

As the closures used to derive Eq.~(\ref{eq_epsilon_HL}), which were modelled on the 
basis of turbulent channel flows and freely decaying turbulence, may not be universal,
they should ideally be obtained from a more general framework. This approach has been taken 
in \citet{canuto10b} who derived a dynamical equation for the TKE dissipation rate $\epsilon$ 
using the general turbulence model of \citet{canuto96d}. That requires the spectrum 
of the source driving turbulence to be known. For shear-driven flows power law spectra for the 
TKE and the Reynolds stress spectrum can readily be specified. Note that these concern scales 
$k < k_0$, i.e., below the maximum of the TKE spectrum $E(k)$. In addition, energy conservation 
is invoked which allows computing the non-local contribution to $\epsilon$ from the flux of turbulent
kinetic energy. That closure was already used in \citet{canuto92b} (Eq.~(37f)) and
the non-local character it introduces into Eq.~(\ref{eq_epsilon}) was discussed in 
Sect.~11 of \citet{canuto93b}. It was tested in \citet{kupka07e} who found it to be one of 
the most robust ones among all the closures suggested for the Reynolds 
stress models of \citet{canuto92b}, \citet{canuto93b}, \citet{canuto98b}, \citet{canuto01b}, and  
in \citet{canuto09b}. It specifies that $\overline{w\epsilon} = (3/2)\tau^{-1}\,F_{\rm kin}$ with 
$F_{\rm kin} = \rho\,\overline{q^2 w}/2$. In practice, the accuracy of this closure is degraded, if 
$\overline{q^2 w}$ can only be computed from a downgradient approximation, but even in
this case it justifies that $D_{\rm f}(\epsilon)$ can be evaluated from
$D_{\rm f}(K)$ which is required anyway. Hence, \citet{canuto10b} use
the (exact) dynamical equation for the turbulent kinetic energy and a closure for $F_{\rm kin}$ to
compute $D_{\rm f}(\epsilon)$. The equivalents of $c_1$ and
$c_2$ of Eq.~(\ref{eq_epsilon_HL}) are obtained from within the model, too. The resulting
dissipation rate equation passes the same tests as the original Eq.~(\ref{eq_epsilon_HL})
for turbulent channel flow and also two tests concerning the shear dominated planetary
boundary of the Earth atmosphere. Unfortunately, this procedure is currently 
not feasible for the case of convection in stars, since this would require accurate knowledge
of the turbulent kinetic energy spectrum over a large range of scales and as a function of depth 
throughout the star (see also the discussions in \citealt{gizon12b} and Fig.~5 in \citet{hanasoge16b}
on difficulties in modelling the turbulent kinetic energy spectrum for the Sun).

The dissipation rate equation Eq.~(\ref{eq_epsilon_HL}) has hence remained part of
the Reynolds stress model of \citet{canuto11b}, whether for dealing with double-diffusive 
convection \citep{canuto11c} or overshooting \citep{canuto11d}. The
latter paper provides a detailed 
discussion of the computation of $\epsilon$, which considers Eq.~(\ref{eq_epsilon_HL}) 
and $\overline{w\epsilon} = (3/2)\tau^{-1}\,F_{\rm kin}$ for non-local contributions. The role of 
gravity waves as a source of dissipation in the overshooting zone is emphasised, too. From earlier 
work of \citet{kumar99b}, it is concluded in \citet{canuto11c} that $\epsilon \approx 10^{-3}\,{\rm cm}^2\,{\rm s}^{-3}$.
However, as also pointed out in \citet{canuto11d}, it is unclear how this result could be
applied to overshooting zones other than the solar tachocline. Thus, in his Eq.~(5h),
\citet{canuto11d} suggests to use $\tau\,\tilde{N} = O(1)$ to compute $\tau$ and hence
via $\tau = 2\,K/\epsilon$ the dissipation rate $\epsilon$ in the overshooting region.
This, however, is consistent with the claim that the term $c_3\, \epsilon\, \tilde{N}$,
neglected in the explicit form of the $\epsilon$-equation in \citet{canuto11b,canuto11c,canuto11d},
actually dominates in the overshooting region.

Recalling \citet{kupka02b} and \citet{montgomery04b}, who had found the term 
$c_3\, \epsilon\, \tilde{N}$ to dominate the solution of Eq.~(\ref{eq_epsilon})
in their applications of the Reynolds stress model of \citet{canuto98b} to overshooting in
envelopes of A-stars and white dwarfs and taking into account the confirmation of their 
results for the case of a DA white dwarf by 3D radiation hydrodynamical simulations in \citet{kupka18b},
 Eq.~(\ref{eq_epsilon}) is still the physically most complete model for the computation
of $\epsilon$ available at the moment. It is thus used to guide the considerations in Sect.~\ref{Sect_GARSTEC}. 
Note that the dynamical equation for $\epsilon$ which is discussed here does not account for physical 
effects due to compressibility. \citet{canuto97b} has presented several different models to extend
Eq.~(\ref{eq_epsilon}) beyond its {\em solenoidal} (incompressible) component $\epsilon_s$ and
account for a dilation (compressible) contribution $\epsilon_d$ (cf.\ Sect.~14 in that paper). For current 
modelling in stellar structure and evolution theory such extensions appear yet too advanced: the very first step
is to give up the MLT approach to compute $\epsilon$ as specified by Eq.~(\ref{eq_MLT_epsilon})--(\ref{eq_MLT_alpha}).

\end{document}